\documentclass[fleqn, usenatbib]{mnras}

\usepackage[T1]{fontenc}

\DeclareRobustCommand{\VAN}[3]{#2}
\let\VANthebibliography\thebibliography
\def\thebibliography{\DeclareRobustCommand{\VAN}[3]{##3}\VANthebibliography}

\usepackage{graphicx}	
\usepackage{amsmath}	
\usepackage{amssymb}	
\usepackage{newtxtext,newtxmath}

\title[Reconstructing burstiness from {[O/Fe]}]{The imprint of bursty star formation on alpha-element abundance patterns in Milky Way-like galaxies}

\author[H. Parul et al.]{
Hanna Parul$^{1}$\thanks{E-mail: hparul@crimson.ua.edu},
Jeremy Bailin$^{1}$, 
Andrew Wetzel$^{2}$, 
Alexander B. Gurvich$^{3}$,
\newauthor 
Claude-Andr{\'e} Faucher-Gigu{\`e}re$^{3}$, 
Zachary Hafen$^{4}$, 
Jonathan Stern$^{5}$, 
Owain Snaith$^{6}$\\
$^{1}$Department of Physics and Astronomy, University of Alabama, Box 870324, Tuscaloosa, AL, 35487, USA\\
$^{2}$Department of Physics \& Astronomy, University of California, Davis, 1 Shields Ave, Davis, CA 95616, USA\\
$^{3}$Department of Physics \& Astronomy and CIERA, Northwestern University, 1800 Sherman Ave, Evanston, IL 60201, USA\\
$^{4}$Department of Physics \& Astronomy, 4129 Reines Hall, University of California, Irvine, CA 92697, USA\\
$^{5}$School of Physics \& Astronomy, Tel Aviv University, Tel Aviv 69978, Israel\\
$^{6}$GEPI, Observatoire de Paris, PSL Research University, CNRS, Place Jules Janssen, 92190, Meudon, France
}

\date{Accepted XXX. Received YYY; in original form ZZZ}

\pubyear{2020}

\begin{document}
\label{firstpage}
\pagerange{\pageref{firstpage}--\pageref{lastpage}}
\maketitle

\begin{abstract}
Milky Way-mass galaxies in the FIRE-2 simulations demonstrate two main modes of star formation. At high redshifts star formation occurs in a series of short and intense bursts, while at low redshifts star formation proceeds at a steady rate with a transition from one mode to another at times ranging from 3 to 7 Gyr ago for different galaxies. We analyse how the mode of star formation affects iron and alpha-element abundance. 
We find that the early bursty regime imprints a measurable pattern in stellar elemental abundances in the form of a "sideways chevron" shape on the [Fe/H] - [O/Fe] plane and the scatter in [O/Fe] at a given stellar age is higher than when a galaxy is in the steady regime.
That suggests that the evolution of [O/Fe] scatter with age provides an estimate of the end of the bursty phase. We investigate the feasibility of observing of this effect by adding mock observational errors to a simulated stellar survey and find that the transition between the bursty and steady phase should be detectable in the Milky Way, although larger observational uncertainties make the transition shallower.
We apply our method to observations of the Milky Way from the Second APOKASC Catalog and estimate that the transition to steady star formation in the Milky Way happened 7-8 Gyrs ago, earlier than transition times measured in the simulations.
\end{abstract}




\section{Introduction}

Star formation in galaxies sometimes has a bursty character, with large fluctuations in the star formation rate (SFR).
In contrast to large nearby spiral galaxies with star formation confined to thin rotationally-supported gas disks, high-redshift galaxies were gas-rich with irregular morphology dominated by star-forming clumps \citep[][]{Elmegreen}.
Scatter in the SFR in individual galaxies in the bursty phase is predicted to be $\sim 0.3$ dex \citep[][]{FloresVelazquez,Gurvich22} comparable to the observed scatter in the population of galaxies on the ``star formation main sequence'' (SFMS) \citep[][]{Orr2017, Sparre, Whitaker2012, Speagle2014, Leja2019}.
One way to test whether star formation is bursty is to study the ratio of SFR indicators sensitive to different timescales. A common choice is to use the H$\alpha$-FUV ratio: H$\alpha$ emission traces star formation over the most recent $\sim$ 5 Myr, while FUV is an indicator of star formation over $\sim 10-100$ Myr; therefore after a short burst of star formation, the ratio would exceed unity at first and rapidly decrease afterwards. This method was used, for example, in \citet{Guo2016} (with H$\beta$-FUV) to reveal bursty star formation in low-mass galaxies at 0.4 < z < 1 and in nearby dwarfs \citep{Weisz2012, Emami2019}. 
Unlike dwarf galaxies which continue to be bursty throughout their life, simulations predict that Milky Way-mass galaxies change their star formation mode from highly variable at early times to calm and steady at later times. 
\citet{Sparre} investigated the scatter in the SFR-$M_{\star}$ relation at z=2 for the FIRE-simulated galaxies and found that due to bursty star formation, the scatter is higher for H$\alpha$-derived SFR than for FUV-based SFR and the difference between the two metrics is consistent with observations. They also showed that the scatter increases with decreasing mass. The response of the SFR indicators to realistic star formation histories in FIRE galaxies was also studied in \citet{FloresVelazquez}. They were aiming to identify the width of the boxcar over which the true SFR should be averaged in order to match the SFR indicated from observations and found that for H$\alpha$ this best-fitting boxcar-averaging timescale is short ($\sim$5 Myr) and relatively stable both in bursty and steady regimes, while for FUV it fluctuates strongly from 10 to 100 Myr during the bursty epoch and stays at the level of 10-20 Myr when SFR is steady.

Variable bursty star formation is considered to be one of the main predictions of simulations with explicitly resolved stellar feedback \citep{FG2018Nature}. \citet{Stinson_2006} used simulations of isolated dwarf galaxies to show that supernova feedback is able to launch periodic episodes of star formation. \citet{Dominguez} studied the effect of bursty star formation in dwarf galaxies from the EAGLE simulations on the ionizing photon production and star formation rate derived from continuum measurement. Multiple studies \citep{Pontzen, Governato, Teyssier} of simulated dwarf galaxies have focused on cusp-flattening processes of dark matter haloes accompanied by the presence of outflows and galactic fountains induced by bursty star formation. Short-scale variability of star formation is also observed in a diverse sample of galaxies from the FIRE simulations \citep{Hopkins_FIRE2, Muratov, AnglesAlcazar_baryon, Sparre, Ma, FG}. Of particular interest are the processes that accompany the transition from bursty to steady regime in Milky Way-mass galaxies. \citet{Muratov} demonstrated that feedback from bursts at early times is able to create "gusty" superwinds and eject a significant mass of dense ISM which leads to temporary suppression of star formation. \citet{Yu} found that the end of the bursty phase correlates with the time when formation of stars with thin-disk kinematics begins to prevail over the formation of stars with thick-disk kinematics; they also found that a more extended steady phase results in a higher thin-disk fraction at z=0. 
\citet{Gurvich22} shows that formation of the thin disk occurs only after the end of the bursty phase. The transition from bursty to steady star formation also correlates with a change in CGM properties from cool ($T\sim10^4$ K) and dynamic to hot ($T\sim10^6$ K) and nearly hydrostatic \citep[][]{Stern}. The connection between thin disk formation and a hot halo may be that the hot halo deposits gas onto the galaxy gently and already-aligned \citep[][]{Hafen2022}.
\citet{Bellardini} reported that a transition to steady star formation is complemented by smoothing of azimuthal abundance variations and establishment of large-scale radial gradients. 
Therefore the transition from bursty to steady mode of star formation marks an important milestone in galactic life, and estimating the transition time based on analysis of observational data can help to test theories of galaxy formation and evolution.

For the Milky Way, the usual method of studying H$\alpha$ and FUV-derived SFR is not suitable for tracing the evolution of burstiness because it provides only the current state of star formation in the galaxy. We also cannot simply count the number of stars as a function of age to resolve individual bursts since uncertainties in age measurements are comparable to or larger than the typical length of a burst.

One of the widely used ways to look into the past star formation history of the Galaxy is to use stellar elemental abundances. 
While high mass stars quickly deposit metals into the interstellar medium in form of stellar winds, mass loss and supernovae explosions, low mass stars show negligible evolution of their surface abundances and preserve the chemical composition of their birthplace environment in their atmospheres.
Different elements are produced on different timescales through various formation channels; therefore the observed patterns of elemental abundance ratios contains the imprint of the star formation history \citep{Tinsley}.

In this paper we use Milky Way-mass galaxies from FIRE-2 to show that the scatter of alpha-elements abundances at a given age is correlated with the burstiness of star formation and can be used to recover the mode of historical star formation and estimate the time of transition from bursty to steady regime. We apply our method to the stellar data from the Second APOKASC Catalog to estimate the end of the bursty period in the Milky Way. 

The paper is organized as follows. In Sect.~\ref{sec:fire2} we describe the set of simulations, in Sect.~\ref{sec:methods} we define burstiness of star formation and investigate how it is manifested through age-abundance relations. We show how to estimate the transition time between bursty and steady modes in Sect.~\ref{sec:results} and apply our method to Milky Way data. We discuss our results in Sect.~\ref{sec:discussion} and conclude in Sect.~\ref{sec:conclusion}.

\section{FIRE-2 simulations}
\label{sec:fire2} 
We use simulations of Milky Way-like galaxies from the FIRE-2 cosmological baryonic zoom-in simulations described in \citet{Hopkins_FIRE2}. The simulations were run using the hydrodynamical code GIZMO \citep[][]{gizmo} in the mesh-free finite-mass Lagrangian Godunov mode. FIRE-2 simulates a number of heating and cooling processes including free-free emission, photoionization/recombination, Compton scattering, photoelectric, metal-line, molecular, fine structure, dust-collisional along with cosmic-ray heating and metallicity-dependent radiative cooling for gas at a range of temperatures $T \sim 10$--${10^{10}}{} ~\mathrm{K}$. The simulations model the cosmic UV background based on \citet{FG2009}. Star formation occurs as described in \citet{Hopkins_FIRE2} in dense ($n > 1000 ~\mathrm{cm^{-3}}$), self-gravitating, Jeans-unstable molecular gas with a local 100\% efficiency per free-fall time. Every star particle is considered as a simple stellar population with Kroupa IMF which evolves along stellar population models from STARBURST99 \citep[][]{starburst99}. Essential for this paper is the detailed simulation of stellar feedback from supernovae of Type Ia (SNIa), core-collapse supernovae (CCSN), and mass loss dominated by winds from O, B and AGB stars. 
The rate of SNe Ia per unit stellar mass per Myr is adopted from \citet{Mannucci_2006} as follows:
\begin{equation}
    \begin{array}{l}
        dN_{\mathrm{Ia}}/dt = 0\mathrm{~for~}t_{\mathrm{Myr}} < 37.53 \\ 
        dN_{\mathrm{Ia}}/dt = 5.3 \times 10^{-8} + 1.6 \times 10^{-5} \exp \{-[(t_{\mathrm{Myr}} - 50) / 10]^2\} \\ \mathrm{~for~}t_{\mathrm{Myr}} \geq 37.53 \\ 
    \end{array}
\end{equation}
These are normalized such that the expectation number of SNe Ia per stellar mass is 0.001$M_{\odot}^{-1}$.
For SNe II the rate is approximated as a piecewise-constant function:
\begin{equation}
    \begin{array}{l}
        dN_{\mathrm{II}}/dt = 0\mathrm{~for~}t_{\mathrm{Myr}} < 3.401\mathrm{~or~} t_{\mathrm{Myr}} > 37.53\\ 
        dN_{\mathrm{Ia}}/dt = 5.408 \times 10^{-4} \mathrm{~for~}3.4 < t_{\mathrm{Myr}} < 10.37 \\ 
        dN_{\mathrm{Ia}}/dt = 2.516 \times 10^{-4} \mathrm{~for~}10.37 < t_{\mathrm{Myr}} < 37.53 \\ 
    \end{array}
\end{equation}
The yields for SN Ia are from \citet{Iwamoto} and for SN II from tables in \citet{Nomoto}
Yields for continuous stellar winds are from the models described in \citet{vandenHoek, Marigo2001, Izzard} and synthesized in \citet{Wiersma}; the mass-loss rate is presented in the IMF-integrated form and has a mild dependence on the metallicity.
The simulations trace 11 elements (H, He, C, N, O, Ne, Mg, Si, S, Ca, Fe), which allows us to constrain the star formation history from the abundance ratio of $\alpha$-elements and iron. 
To account for enrichment of gas particles from neighboring particles through the unresolved metal diffusion (e.g. Rayleigh-Taylor and Kelvin-Helmholtz instabilities and turbulent eddies) the simulations include a model examined in \citet[][]{Shen2010}:
\begin{equation}
\begin{array}{c}
\displaystyle\frac{\partial \mathrm{M}_i}{\partial t}+\nabla \cdot\left(D \nabla \mathrm{M}_i\right)=0,\\ 
    D=C_0\|\mathrm{\mathbf{S}}\|{ }_f \mathrm{\mathbf{h}}^2, 
\end{array}
\end{equation}
where $\mathrm{\mathbf{h}}$ is the resolution scale, $\mathrm{\mathbf{S}}$ is the shear tensor, and $C_0$ is a constant calibrated from numerical simulations \citep[][]{Su2017}.

In this paper we used 11 galaxies --- 5 galaxies from the Latte suite \citep{Wetzel_latte, refA, Hopkins_FIRE2, refD} which are broadly consistent with properties of M31 or the Milky Way and 3 Local Group-like pairs from ``ELVIS on FIRE'' \citep{refD, refE}. We limited our analysis to in-situ stars i.e. those that formed within 20 kpc comoving of the host center. Properties of the simulated galaxies are listed in Table~\ref{tab:galaxies}.

\begin{table*}
    \centering
\begin{tabular}{c|c|c|c|c|c|c}
     Simulation & $M_{\star}$ & $M_{\mathrm{halo}}$ & $m_i$ & $t_{\mathrm{B, ~[O/Fe]}}$ & $t_{\mathrm{B, ~burst}}$ & Reference\\
     Name & $[\mathrm{M_{\odot}}]$ & $[\mathrm{M_{\odot}}]$ & $[\mathrm{M_{\odot}}]$ & [Gyr] & [Gyr] \\
     \hline
     m12f & $8.8\times10^{10}$ & $1.4\times10^{12}$ & 7070 & 4.71 & 4.98 & A\\
     m12i & $7.0\times10^{10}$ & $9.7\times10^{11}$ & 7070 & 3.63 & 3.21 & B\\
     m12m & $1.3\times10^{11}$ & $1.9\times10^{12}$ & 7070 & 4.35 & 3.79 & C\\
     m12c & $6.4\times10^{10}$ & $1.9\times10^{12}$ & 7070 & 3.26 & 3.70 & D \\
     Thelma & $7.9\times10^{10}$ & $3.4\times10^{12}$ & 4000 & 3.26 & 2.57 & D \\
     Louise & $2.9\times10^{10}$ & $1.4\times10^{12}$ & 4000 & 5.08 & 5.53 & D \\
     Romeo & $7.4\times10^{10}$ & $1.1\times10^{12}$ & 3500 & 6.53 & 6.62 & D\\
     Juliet & $4.2\times10^{10}$ & $9.1\times10^{11}$ & 3500 & 4.36 & 4.48 & D\\
     m12b & $9.4\times10^{10}$ & $1.2\times10^{12}$ & 7070 & 7.26 & 6.32 & D\\
     Romulus & $1.0\times10^{11}$ & $1.7\times10^{12}$ & 4000 & 5.44 & 4.88 & E\\
     Remus & $5.1\times10^{10}$ & $1.1\times10^{12}$ & 4000 & 5.44 & 5.83 & E\\
\end{tabular}
    \caption[caption]{Properties of galaxies from FIRE-2 simulations used in this paper. We provide the stellar mass ($M_{\star}$) within the central 20 kpc comoving at z = 0, the halo virial mass ($M_{\mathrm{halo}}$) defined by $\rho_{200\mathrm{m}}$, the mass resolution for baryonic particles ($m_i$) and transition times derived from analysis of scatter of [O/Fe] as a function of age ($t_{\mathrm{B, ~[O/Fe]}}$) and from analysis of burstiness ($t_{\mathrm{B, ~burst}}$) following the definition in \protect\citet{Yu}.\\\hspace{\textwidth}
    The references are: A: \protect\citet{refA}, B: \protect\citet{Wetzel_latte}, C: \protect\citet{Hopkins_FIRE2}, D: \protect\citet{refD}, E: \protect\citet{refE}}
    \label{tab:galaxies}
\end{table*}


\section{Analysis}
\label{sec:methods}
\subsection{Star Formation Histories}
\label{sfh}
We begin with characterizing the mode of star formation as a function of time using the example of galaxy \texttt{m12i}. We present results for all 11 simulations in Appendix \ref{sec:appA}. The purple line on Fig.~\ref{fig:std_11} shows the instantaneous SFR (measured in 1 Myr bins) normalized by the SFR averaged over a 500 Myr rolling window. There are two distinct regimes in which star formation occurs. At early times ($\displaystyle t_\mathrm{lookback} \gtrsim$ 3-4 Gyr) star formation was ``bursty'' and characterized by high variability with 10-40 Myr timescale. Recently, over the past 3-4 Gyr, star formation proceeded in a ``steady'' mode with relatively constant star formation rate. 
 Fig.~\ref{fig:ofe_zoom} gives a closer look at these two modes --- in the ``bursty'' regime, star formation (blue line) on the top panel occurs in bursts with width $\approx 40-50$ Myr and intensity up to 10 times higher than the average value, while the bottom panel shows ``steady'' regime variations  which are much shorter and do not exceed the mean level of star formation rate by much.

\begin{figure*}
	\includegraphics[width=2.\columnwidth]{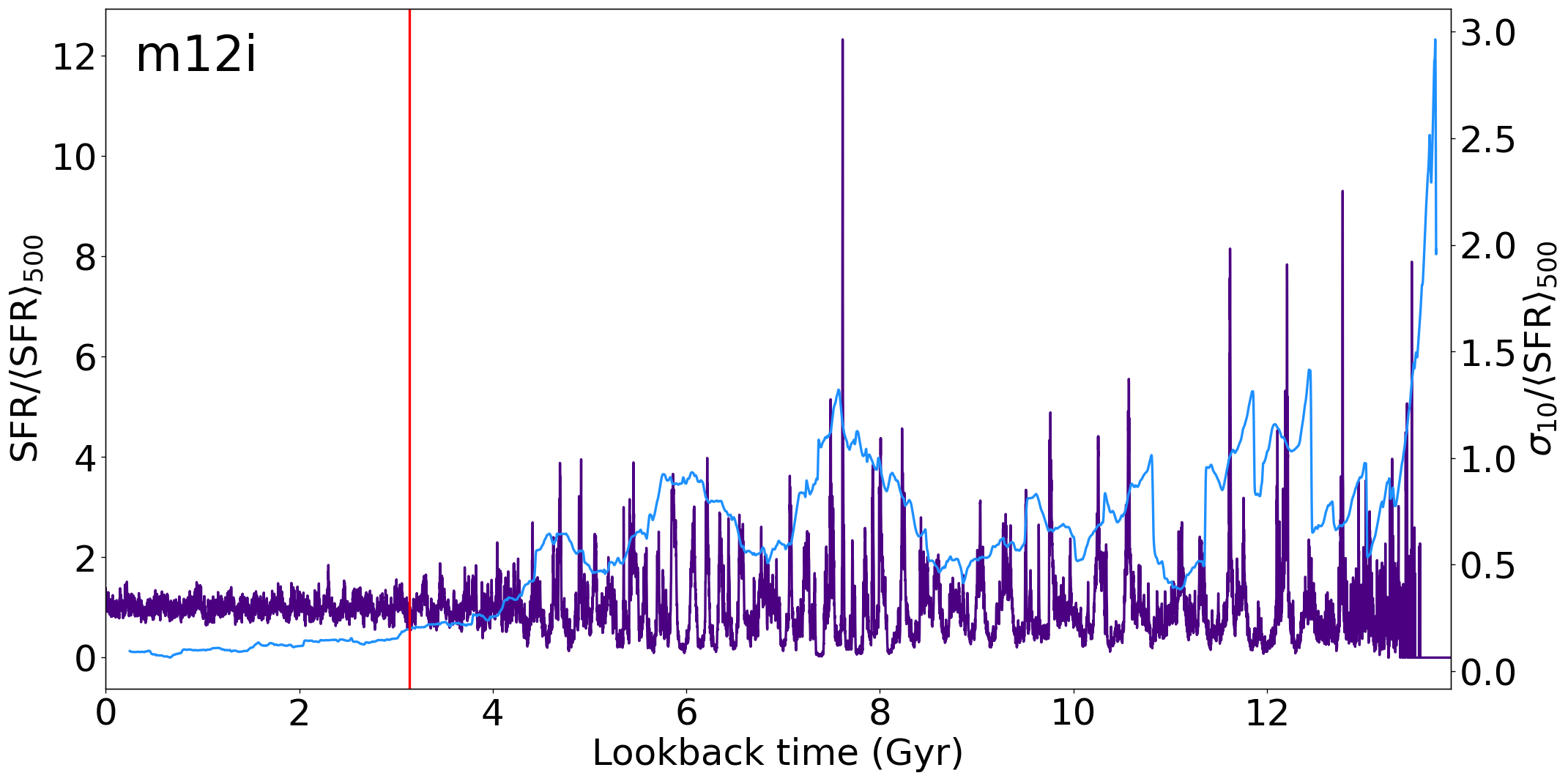}
    \caption{Normalized SFR (purple line) overplotted with burstiness, $\sigma_{10}/\mathrm{SFR}$ (blue line). The vertical line corresponds to the time when $\sigma_{10} / \left<SFR\right>_{500}$ drops below 0.2 which separates bursty and steady modes.}
    \label{fig:std_11}
\end{figure*}

To quantify the variability of star formation and define the time of transition between modes we follow the definition of burstiness in \citet{Yu}: burstiness is computed as the ratio of the standard deviation of star formation measured in 10 Myr bins to the star formation averaged over 500 Myr, $\sigma_{10}/\left<SFR \right>_{500}$ and is shown with a blue line on Fig.~\ref{fig:std_11}. The red vertical line on Fig.~\ref{fig:std_11} marks the end of the bursty phase, $t_{\mathrm{B, ~burst}}$, which is defined as time when burstiness falls below 0.2:

\begin{equation} \label{eq:1}
\frac{\sigma_{10}(t_{\mathrm{B, ~burst}})}{\left<SFR \right>_{500}(t_{\mathrm{B, ~burst}})} = 0.2.
\end{equation}

\begin{figure*}
    \centering
    \includegraphics[width=2\columnwidth]{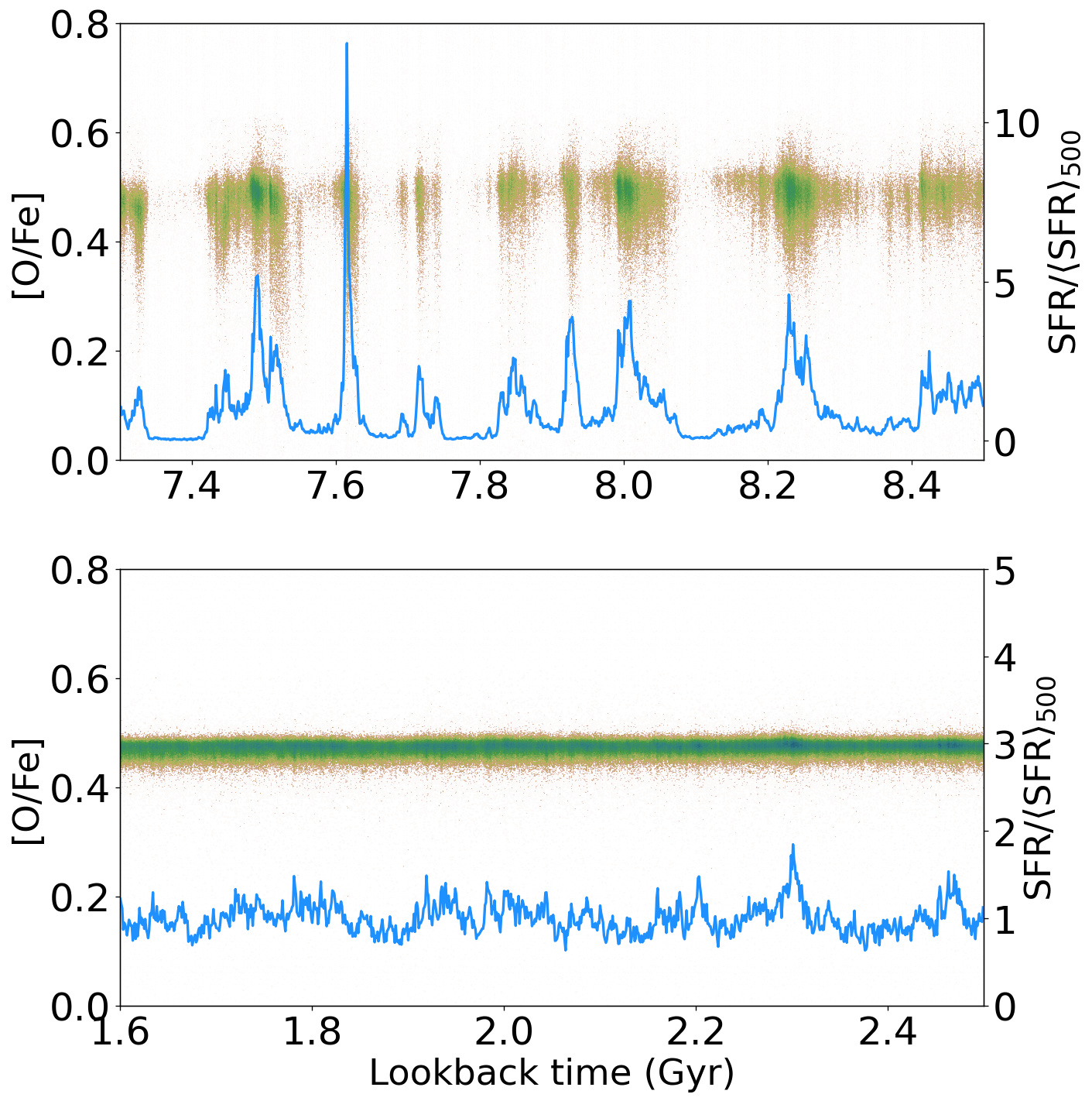}
    \caption{Evolution of [O/Fe] for a short interval during the bursty (\textit{top}) and steady (\textit{bottom}) modes of star formation. The blue line shows normalized star formation rate $\mathrm{SFR/\left<SFR\right>_{500}}$ for galaxy \texttt{m12i}. Peaks of star formation corresponds to a wider range of [O/Fe].}
    \label{fig:ofe_zoom}
\end{figure*}

The burstiness for all galaxies, $\sigma_{10}/\mathrm{\left<SFR\right>_{500}}$ (red line), with vertical lines indicating the end of the bursty regime are shown on panels of 
Fig.~\ref{fig:std_all} 
and estimated transition times are reported in Table \ref{tab:galaxies}.

\subsection{Age-abundance relation}
The distribution of stellar ages and abundance ratios such as [O/Fe] provides a way to infer historical star formation since stars of different masses inject different elements into the ISM on a different timescales. The amount of oxygen in the galaxy increases due to core-collapse supernova (CCSN) explosions that occur on a relatively short timescales while iron formation is dominated by type Ia supernovae that need longer to evolve and explode (in the FIRE-2 implementation SNIae begin when the age of a star particle $t > 37.53$ Myr). 
Therefore episodes of active star formation are characterized by high [O/Fe] values and the transition to moderate star formation is accompanied by a decrease of [O/Fe] and an increase in iron content.

\begin{figure*}
    \centering
    \includegraphics[width=2\columnwidth]{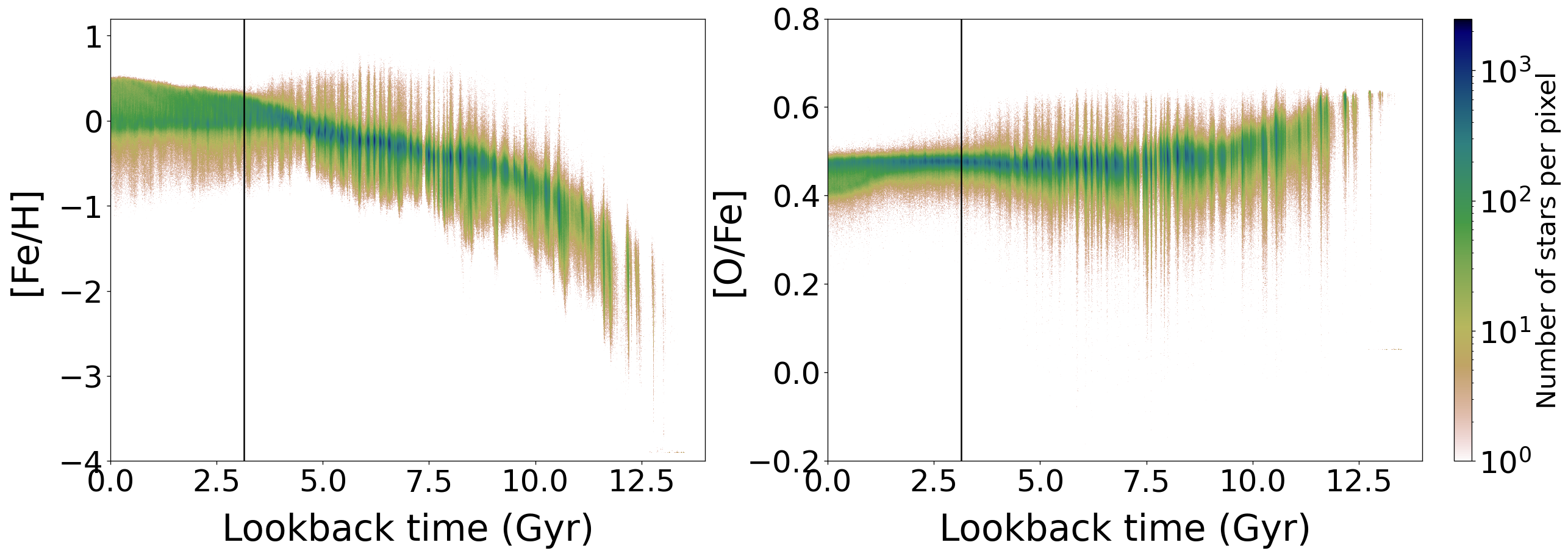}
\caption{Simulation \texttt{m12i}: evolution of [Fe/H] and [O/Fe], our tracer of [$\alpha$/Fe], with lookback time. The vertical line marks the end of the bursty phase. The distribution of [O/Fe] during the bursty regime (to the right of the vertical line) has a noticeably wider scatter.}
\label{fig:age_feh_age_ofe} 
\end{figure*}

We now consider the elemental abundance ratio and metallicity for stars of different ages in the galaxy \texttt{m12i} as an example (see Appendix~\ref{sec:appA} for details on the other galaxies). Fig.~\ref{fig:age_feh_age_ofe} shows stellar metallicity [Fe/H] as a function of lookback time on the left panel and relative oxygen abundance [O/Fe] also as a function of lookback time on the right panel. The iron content in the galaxy increases with time with younger stars being more metal-rich while relative oxygen ratio stays nearly constant with minor decrease at low redshift. 
It should be noted here, that the absolute values of [O/Fe] are larger than those typically observed in the Milky Way and similar galaxies. 
This difference is likely due to yields and delay time distribution (DTD) of SN Ia implemented in FIRE-2. \citet{Gandhi2022} showed that the values of [Fe/H] increase for the runs with an updated DTD model from \citet{maoz} due to boost in number of SN Ia events but stay in agreement with observational constraints (see Fig. 8 in \citet{Gandhi2022}). In combination with possible changes in star formation history this could potentially bring down [O/Fe] to near-solar values without significantly affecting [Fe/H].

The most glaring feature of the time evolution of the relative oxygen abundance is the tightening of the distribution at times smaller than $\approx 3$ Gyr, to the left of the vertical line which marks the end of the bursty phase (\ref{eq:1}). The width of the [Fe/H] distribution increases in the steady phase which is related to the development of radial metallicity gradients in the disk \citep[][]{Bellardini}.

In Fig.~\ref{fig:ofe_zoom} we zoom in on a short interval during the bursty (top panel) and steady (bottom panel) periods and plot both the normalized SFR and [O/Fe]-age distributions. This direct comparison reveals a perfect match between regions with wider distribution on the age-[O/Fe] plane and individual bursts of star formation at earlier lookback times.

\begin{figure*}
    \centering
    \includegraphics[width=2\columnwidth]{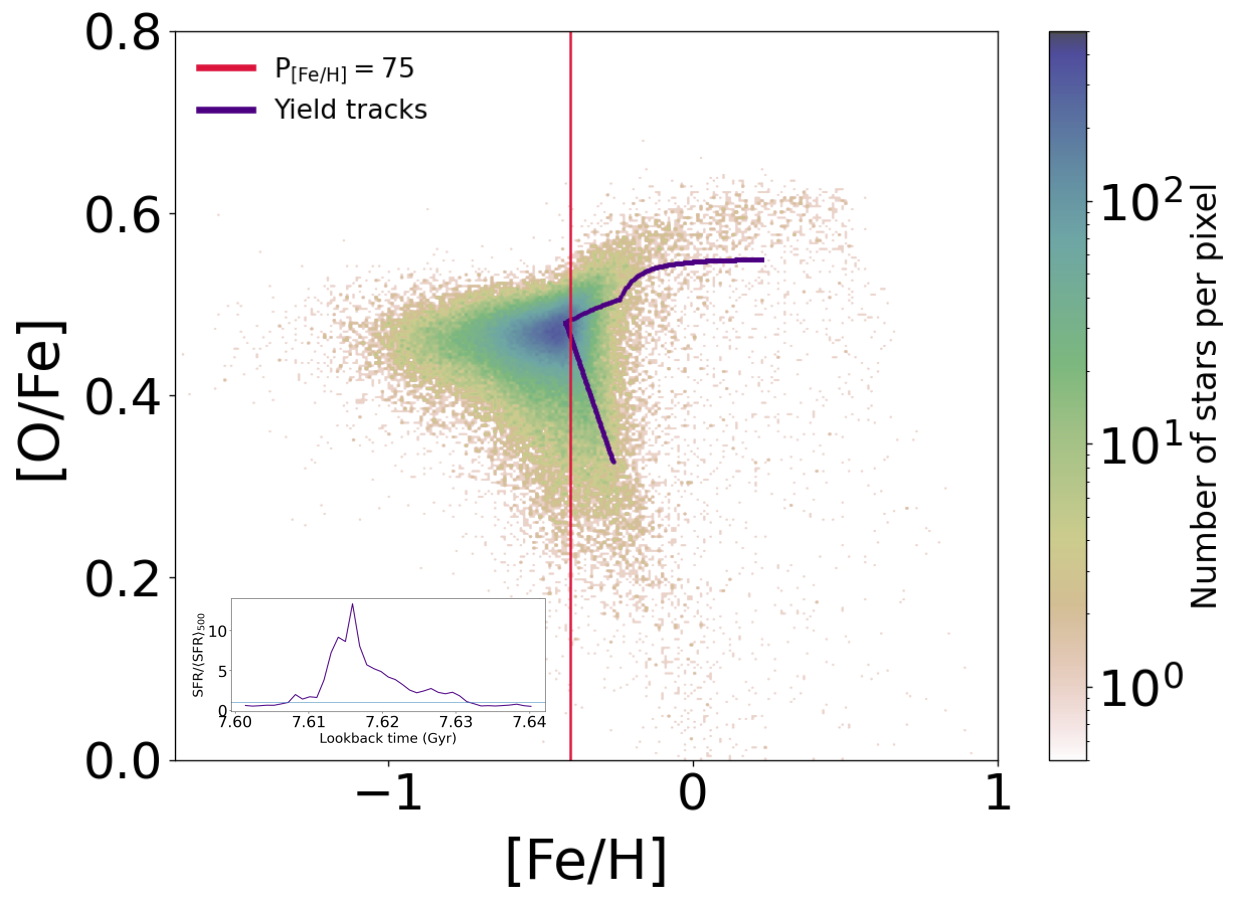}
    \caption{[Fe/H]-[O/Fe] diagram for the simulation \texttt{m12i} plotted for stars born during a short burst between 7.6 and 7.64 Gyr ago (shown on the subplot). Colour codes the density of datapoints in each pixel. The 2D distribution has a clear sideways chevron shape: the arms of the chevron follow yield tracks (purple line) calculated for enrichment solely from SNIa (lower arm) and SNII and OB/AGB stars (upper arm). The vertical line indicates the metallicity corresponding to the 75th percentile of stars born during this time interval.} 
    \label{fig:yields}
\end{figure*}

The distribution of stars on the [Fe/H]-[O/Fe] diagram also behaves differently in steady and bursty regimes. In general, older populations born during the bursty regime, have a wide range of [O/Fe], while younger stars are concentrated along a narrow sequence.
An interesting feature of the bursty mode is that the stars are arranged in a ``sideways chevron'' shape with two tails spreading into areas of lower and higher [O/Fe]. An example of such a distribution is shown on Fig.~\ref{fig:yields} for stars born during the burst of star formation which happened 7.6 - 7.64 Gyr ago. Purple lines outlining the ``chevron'' show the yield tracks for CCSN and AGB stars (top arm) and SNIa stars (lower arm). To compute yield tracks we started with $\mathrm{[Fe/H]_{init}} = -0.55$ and $\mathrm{[O/Fe]_{init}} = 0.48$, corresponding to the iron and relative oxygen abundance of the majority of the stars born during the burst, then we used the rate of SNe Ia, CCSN and mass-loss rate of OB/AGB stars along with ejecta yield masses to obtain the change in oxygen and iron content as a function of time (fits to the stellar evolution models and yields are described in App. A \citet{Hopkins_FIRE2}).

A consistent explanation for the presence of a ``chevron'' shape in the [Fe/H]-[O/Fe] plane is as follows: an episode of intense star formation leads to rapid enrichment of the ISM with a significant amount of alpha-elements. This is responsible for the upper arm of the chevron, as confirmed by the fact that this arm is aligned with the yield track for CCSN and AGB stars (upper purple line on Fig.~\ref{fig:yields}). 
However, in the bursty mode, new stars are born mostly in a small number of dense clumps \citep{Oklopcic, Sparre, FG} therefore strong stellar feedback following the burst of star formation leads to the generation of outflows, quick depletion of gas, and subsequent cessation of star formation. Gas from the outflows mixes with  pristine gas in the halo and falls back down onto the galactic disk providing fuel for a new episode of star formation \citep{AnglesAlcazar_baryon} but with lower metallicity than the metallicity at the end of chevron arm. The lower arm of the chevron is inhabited by stars formed from the gas enriched by SN Ia explosions which are the main source of iron in the ISM; therefore the lower arm on the [Fe/H]-[O/Fe] diagram is aligned with yield vector from SN Ia (Fig.~\ref{fig:yields}). The delay of SN Ia explosions relative to the start of star formation, along with the turbulent and ``patchy'' \citep{HaywardHopkins} character of the ISM during the bursty epoch prevents effective mixing of chemical elements in the ISM during each burst and keeps the chevron arms separated. This is also in agreement with high azimuthal variation of elemental abundances in simulated gas disks at high redshifts explored by \citet{Bellardini}. In the regime with lower instantaneous star formation rate, metals produced by each CCSN or SNIa event have enough time to mix with the surrounding gas before new star particles are formed, unlike in the bursty regime when star formation occurs on timescales shorter than the local mixing time. A similar effect was observed in dwarf galaxies in \citet{patel2021}, for the case of \texttt{m11b} galaxy they show diagonal stripes on the [Fe/H] - [Mg/Fe] plane resulting from individual bursts of star formation. Unlike in dwarf galaxies from \citet{patel2021} we do not see strong age gradient along arms of the chevron because complex temporal and spatial structure of more massive galaxies makes it more difficult to pick out the age trend.

In summary: during the bursty regime, the age-[O/Fe] distribution is wider than during the steady regime, and stars on the [Fe/H]-[O/Fe] plane are organized in a chevron shape. These results suggest that the width of the [O/Fe] distribution measured at a particular time might be a good proxy for the regime in which star formation occurs.

\subsection{Measuring the scatter of the age-[O/Fe] distribution}

To quantify the degree of scatter, referred to as $w_{84}$, we computed the difference between the 8th and 92nd percentiles of [O/Fe] for stars born within a certain time bin. 
The width of the bin, $\delta t$ depends on the time scale of variablility that we want to capture. 
We tested bin widths of 70, 230, 400 and 900 Myr and found that the overall shape of the curve remains the same for different $\delta t$, although shorter bin width leads to a noisier measurement, especially at early times when each bin has relatively few stars with highly varied abundances. Therefore for the rest of the paper we will use $\delta t = 230$ Myr as it provides sufficient time resolution without information loss.

To increase the contrast between $w_{84}$ values measured in bursty and steady modes we filtered stars based on their metallicity [Fe/H]. Fig.~\ref{fig:yields} shows the [Fe/H] - [O/Fe] distribution of stars selected during a bursty phase within a narrow range of ages from 7.6 to 7.64 Gyr, with the vertical line indicating the 75th percentile on [Fe/H] for galaxy \texttt{m12i}. The distribution of the metal-rich stars (those to the right of the vertical line)  born at this time interval are dominated by the open arms of the chevron, rather than by the overdensity at and to the left of its vertex, and therefore have a wider [O/Fe] distribution.

\begin{figure*}
	\includegraphics[width=2\columnwidth]{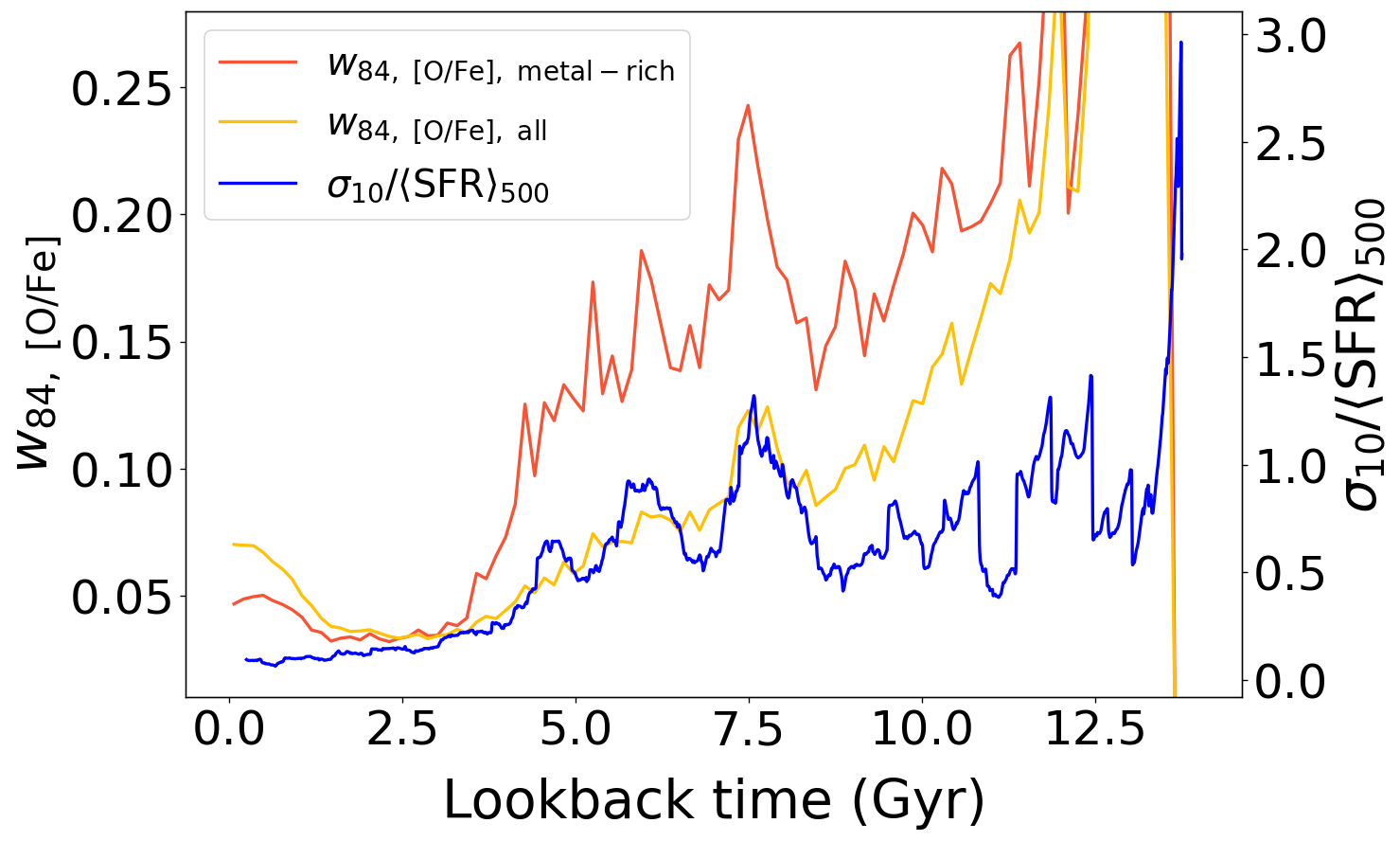}
    \caption{Burstiness, $\sigma_{\mathrm{SFR}}$ (blue curve), and the width of the [O/Fe] distribution, $w_{84}$, as a function of time, measured for stars above the 75th percentile of [Fe/H] at a given age (red curve), vs. for all stars (yellow curve). The scatter is in excellent agreement with the burstiness, however $w_{84}$ measured for high-metallicity stars shows a larger difference between values in bursty and steady mode.}
    \label{fig:compare_high_fe}
\end{figure*}

Fig.~\ref{fig:compare_high_fe} shows the effect of the applied metallicity cut on the scatter, $w_{84, \mathrm{[O/Fe]}}$. Blue line shows the burstiness, $\sigma_{10}/\mathrm{SFR}_{500}$, yellow line -- $w_{84, ~\mathrm{[O/Fe]}}$ measured for all stars, and the red line shows the evolution of $w_{84, ~\mathrm{[O/Fe]}}$ when only metal-rich stars at given age were considered. The first thing to notice is that all three curves show the same trend and transition to a plateau at the similar time, around 3 Gyr ago, together with the end of bursty phase of star formation marked by a vertical line. However the measurement for metal-rich stars demonstrates a larger contrast between values measured in bursty and steady modes, providing an easily detectable signal for the end of the bursty regime.

\subsection{Observational uncertainties}\label{uncert}

With realistic observational uncertainties, features in age-abundance space can get washed out \citep[e.g.][]{Snaith2016, patel2021} which could potentially lead to a flatter shape of $w_{84}$ and complicate the estimation of the transition time. To estimate the effect of uncertainties we used following simple model.

\begin{figure*}
	\includegraphics[width=2\columnwidth]{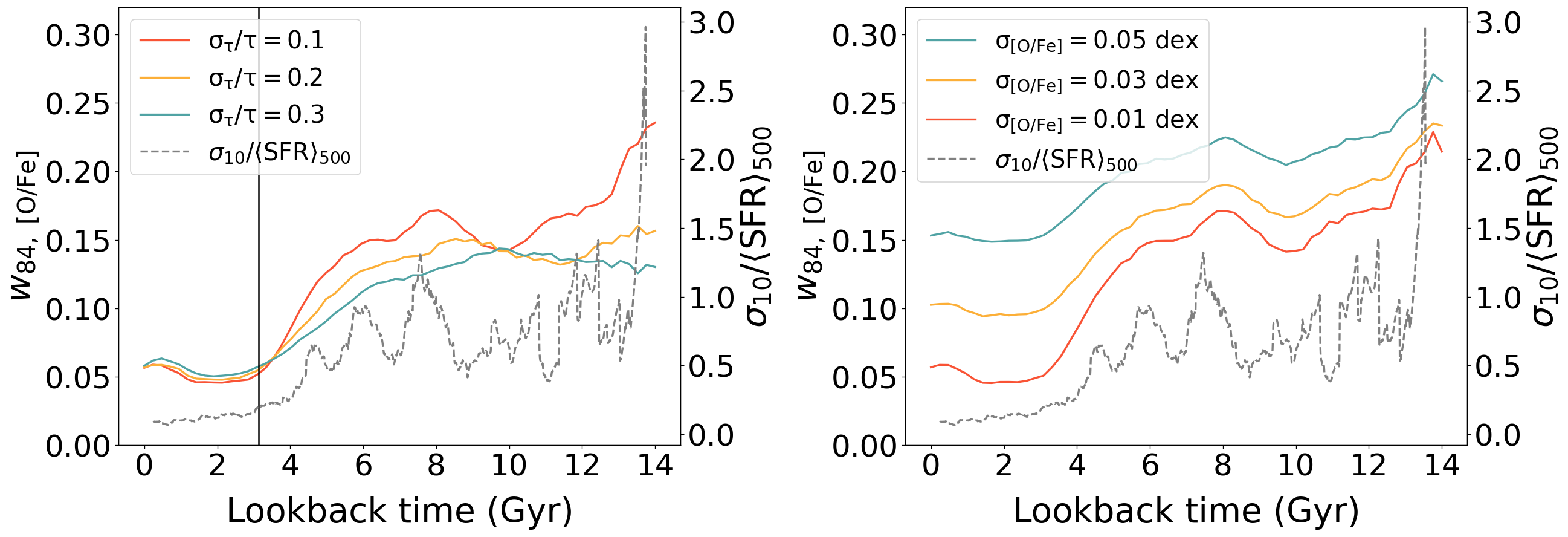}
    \caption{Comparison of $w_{84}$ for data with different implemented error models for the case of galaxy \texttt{m12i}: \textit{left:} age uncertainties of $\sigma_{\tau}/\tau = \{0.1, 0.2, 0.3\}$ and abundance uncertainty 0.01 dex, \textit{right:} abundance uncertainty $\sigma_{\mathrm{[O/Fe]}} = \{0.01, 0.03, 0.05\}$ dex and age uncertainties of $\sigma_{\tau}/\tau = 0.1$.}
    \label{fig:all_err}
\end{figure*}

For age uncertainties, we drew values from a normal distribution with standard deviation $\sigma_{\tau}$ where $\sigma_{\tau}/\tau$ = {0.1, 0.2, 0.3} and $\tau$ is the age of the star. The chosen values were motivated by output uncertainties for ages provided in \citet{3mlnstars}, where ages, masses, and distances were determined through the Bayesian approach based on the astrometric, photometric, and spectroscopic data.
The resulting $w_{84}$ measurements are shown on Fig. \ref{fig:all_err} (left panel). Larger age uncertainties (blue line) make the transition shallower because of the redistribution of stars formed during the bursts of star formation into adjacent age bins. However even with 50\% age errors it is still possible to recover the transition time to the steady regime as a drop in $w_{84}$. 

For abundances, we explored uncertainties of 0.01, 0.03 and 0.05 dex. The range of values for uncertainties was picked based on the median uncertainties for stars from the APOKASC-2 Catalog \citep[][]{apokasc}: $\sigma_{\mathrm{[Fe/H], ~med}} = 0.03$ and  $\sigma_{\mathrm{[\alpha/Fe], ~med}} = 0.01$. Fig.~\ref{fig:all_err} (right panel) shows the effect of increased abundance error. The presence of larger observational errors increases the level of $w_{84}$ which is especially noticeable in the steady regime (lookback time < 3.5 Gyr) and suppresses the amplitude of variation; however a rise beyond $\approx 3-4$ Gyr is still apparent in all cases. 

Another possible source of discrepancy between observations and simulation is the size of the sample. Small numbers of stars in a bin can introduce artificial increases to $w_{84}$ due to poor statistics. We tested samples containing 10\% and 1\% of stars randomly chosen from the initial sample, which resulted in $1.1\times10^6$ and $1.1\times10^5$ star particles respectively.
We found that reduced sample size does not change the overall shape of the curve, but amplifies the short-scale variations and increases the error of the width measurement. 

To explore the effect of the selection function on our measurement, we used mock surveys of synthetic stars generated with the Ananke framework \citep[][]{Sanderson} where each simulation's star particle represented a single stellar population. The survey was designed to resemble the Gaia DR2 catalog in data structure, magnitude limits and observational errors with a self-consistent dust extinction map, computed directly from the gas metallicity distribution in the simulation. We used surveys for galaxies \texttt{m12i}, \texttt{m12f} and \texttt{m12m} presented in \citet{Sanderson} for three solar viewpoints resulting in 9 surveys in total. For ages and abundances of stars in these surveys we applied the same simple error model as for star particles from the simulations.

\begin{figure}
	\includegraphics[width=\columnwidth]{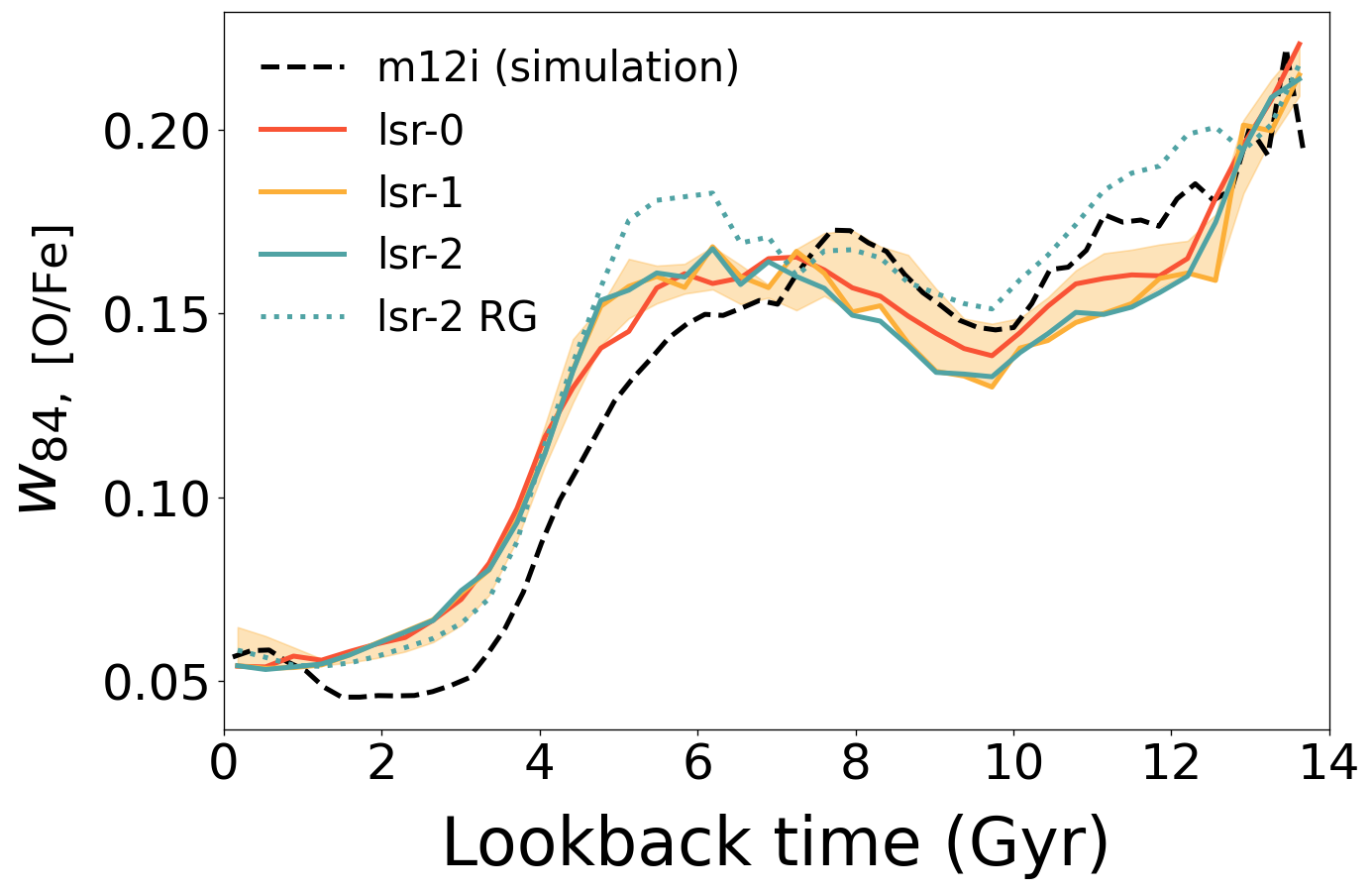}
    \caption{$w_{84}$ as a function of time measured for stars from the simulation \textit{(dashed line)} and three synthetic surveys \textit{(solid lines)} generated with Ananke framework \protect\citep[][]{Sanderson}. Shaded region shows the variation in measured $w_{84}$ due to the different survey depth. Both simulations and surveys include error model with $\sigma_{\tau}/\tau = 0.1$ and $\sigma_{\mathrm{[O/Fe]}} = \sigma_{\mathrm{[Fe/H]}} = 0.01$ dex. The dotted line shows $w_{84}$ for a subset of evolved stars selected such that their temperature and surface gravity match the properties of stars in the APOKASC-2 catalog: $3800~\mathrm{K} <T_{\mathrm{eff}} < 5400~\mathrm{K}$ and $1.0 < \log g < 3.5$. The similarity of all functions means that selection effects in observations should not limit the ability to recover the transition time from bursty to steady star formation.}
    \label{fig:mocks_age_err}
\end{figure}

Fig.~\ref{fig:mocks_age_err} compares $w_{84}$ measured in synthetic surveys (solid lines) to the $w_{84}$ in the simulation (dashed line). To the star particles both in simulation and surveys we applied an error model with following uncertainties: $\sigma_{\tau}/\tau = 0.1$ and $\sigma_{\mathrm{[O/Fe]}} = \sigma_{\mathrm{[Fe/H]}} = 0.01 $ dex.
Each solid line corresponds to the measurement made for stars from the 3 kpc region centered at three different solar viewpoints which have the same galactocentric distance ($R_{\odot} = 8.2$ kpc) but different azimuthal angle. The shaded region shows the variation in $w_{84}$ measured for different survey with depths ranging from 3 kpc to 15 kpc. In order to make a proper comparison with the observational data (e.g. APOKASC-2 Catalog \citep[][]{apokasc}) we restricted our analysis to subsample of synthetic stars with the same temperature and surface gravtiy as the evolved stars in the APOKASC-2: $3800~\mathrm{K} <T_{\mathrm{eff}} < 5400~\mathrm{K}$ and $1.0 < \log g < 3.5$ (dotted line on the Fig.~\ref{fig:mocks_age_err}) The widths calculated from the synthetic surveys are in excellent agreement with those measured from the simulation despite the fact that stellar age distribution ``observed'' in the survey is dominated by young stars. This means that the correlation between the width of the [O/Fe] distribution and burstiness could be in principle observed in the Milky Way regardless of the presence of selection effects.

\subsection{Transition times} 
\label{sec:results}
As we showed in Fig.~\ref{fig:compare_high_fe}, the transition to the steady regime manifests itself as a drop in burstiness, and is associated with a similar drop in $w_{84}$. One interesting feature of the left panel on Fig.~\ref{fig:all_err}, which investigates the effect of age uncertainties on $w_{84}$, is that all three lines converge at the time close to the end of the bursty phase marked by the vertical line. That suggests that we can calculate the transition time based on $w_{84}$ by convolving the data with different age uncertainties and finding the latest time at which they agree.

\begin{figure}
	\includegraphics[width=\columnwidth]{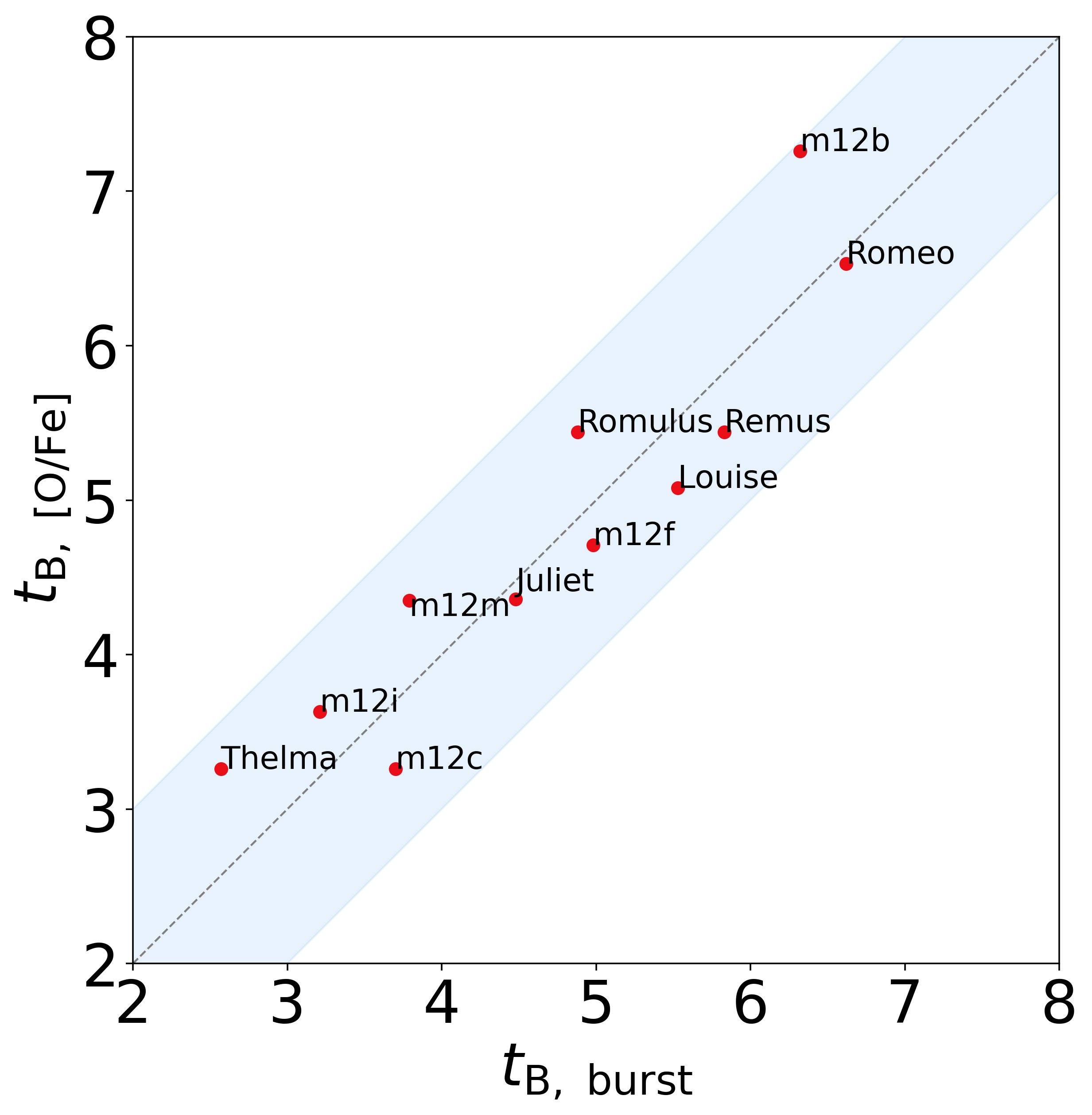}
    \caption{Transition times between the bursty and steady mode of star formation for all 11 galaxies. The transition time measured from burstiness analysis is on the x-axis, and the y-axis contains the transition time from analysis of [O/Fe] scatter with following error model applied: $\sigma_{\tau}/\tau = 0.1$ and $\sigma_{\mathrm{[O/Fe]}} = \sigma_{\mathrm{[Fe/H]}} = 0.01$ dex. The dashed line shows the 1:1 relation; all galaxies lie close to the 1:1 line with deviations less than 1 Gyr (shown with shaded region). }
    \label{fig:time}
\end{figure}

Fig.~\ref{fig:time} summarizes the estimates of the transition times for all simulations in the sample. 
Here the x-axis is the transition time measured from burstiness analysis, while the y-axis contains the transition time estimated from the width measurement. 
All galaxies lie close to the 1:1 line with a deviation of 1 Gyr at most between $t_{w_{84}}$ and $t_{\sigma_{\mathrm{SFR}}}$, meaning that analysis of the scatter of [O/Fe], $w_{84}$, for simulated galaxies provides a good estimate of the end of the bursty epoch.

\subsection{Milky Way analysis}

In previous sections we showed that our method of inferring the mode of star formation is robust in the presence of realistic age and abundance uncertainties, sample size, and probed volume. We now apply it to observational data to estimate the end of the bursty epoch for the Milky Way.

The observational data sets we used are the Second APOKASC Catalog \citep{apokasc} containing stellar properties for 6676 evolved stars with spectroscopic data that were obtained as a part of the APOGEE project \citep{apogee} and asteroseismic data from the  \textit{Kepler} mission \citep{kepler}.
The catalog includes evolutionary state, surface gravity, mean density, mass, radius, along with the age, total metallicity, [M/H], and alpha-elements abundance ratio, $\mathrm{[\alpha/Fe]}$, that is required for our analysis.

\begin{figure*}
	\includegraphics[width=2\columnwidth]{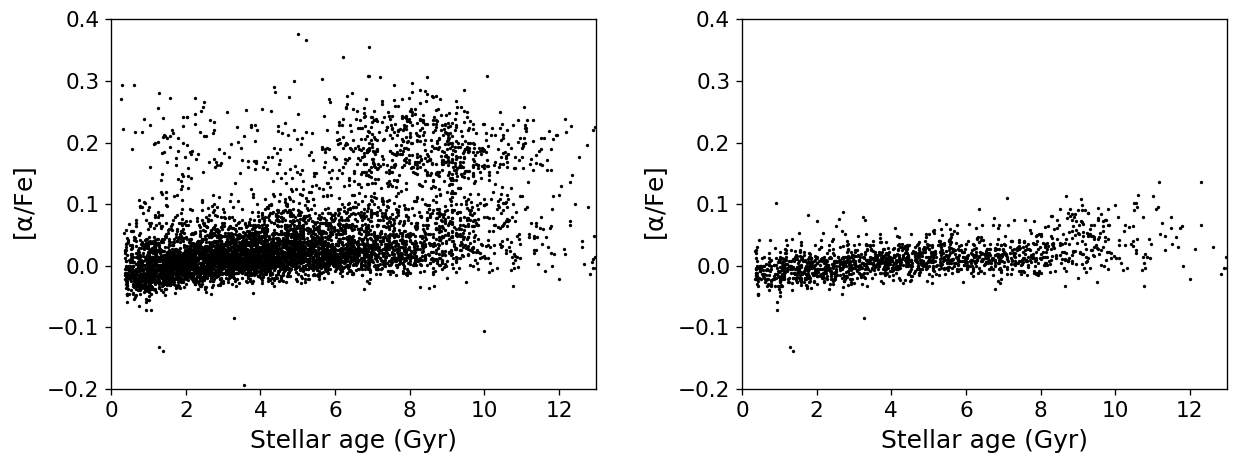}
    \caption{$\mathrm{[\alpha/Fe]}$ as a function of lookback time for stars from Second APOKASC catalog \textit{(left)} and the high-metallicity subsample \textit{(right)}. The high-[$\alpha$/Fe] sequence, which is present in the full sample, disappears for stars filtered by their metallicity and, therefore, does not dominate the measurement of $w_{84}$.}
    \label{fig:MW_highFe}
\end{figure*}

Left panel of Fig.~\ref{fig:MW_highFe} shows that the distribution of APOKASC-2 stars on the age-[O/Fe] plane has a prominent bimodality at earlier times (for stars older than 6 Gyr). Part of this bimodality is explained by the metal-poor thin disk stars that were the subject of discussion in several works \citep{Haywood2008_mptd, Haywood2013_mptd}; it is suggested that these stars formed early in the outer parts of the thin disk and migrated to the solar neighbourhood. While they are not related to the bursty star formation, one might be concerned that this bimodality would increase the scatter of [O/Fe] at early times and mimic the signal that we associate with burstiness. However the right panel of Fig.~\ref{fig:MW_highFe} demonstrates that the higher sequence disappears when only metal-rich stars at given age are considered; therefore our measurement of $w_{84}$ measurement would not be dominated by the bimodality.

\begin{figure}
	\includegraphics[width=\columnwidth]{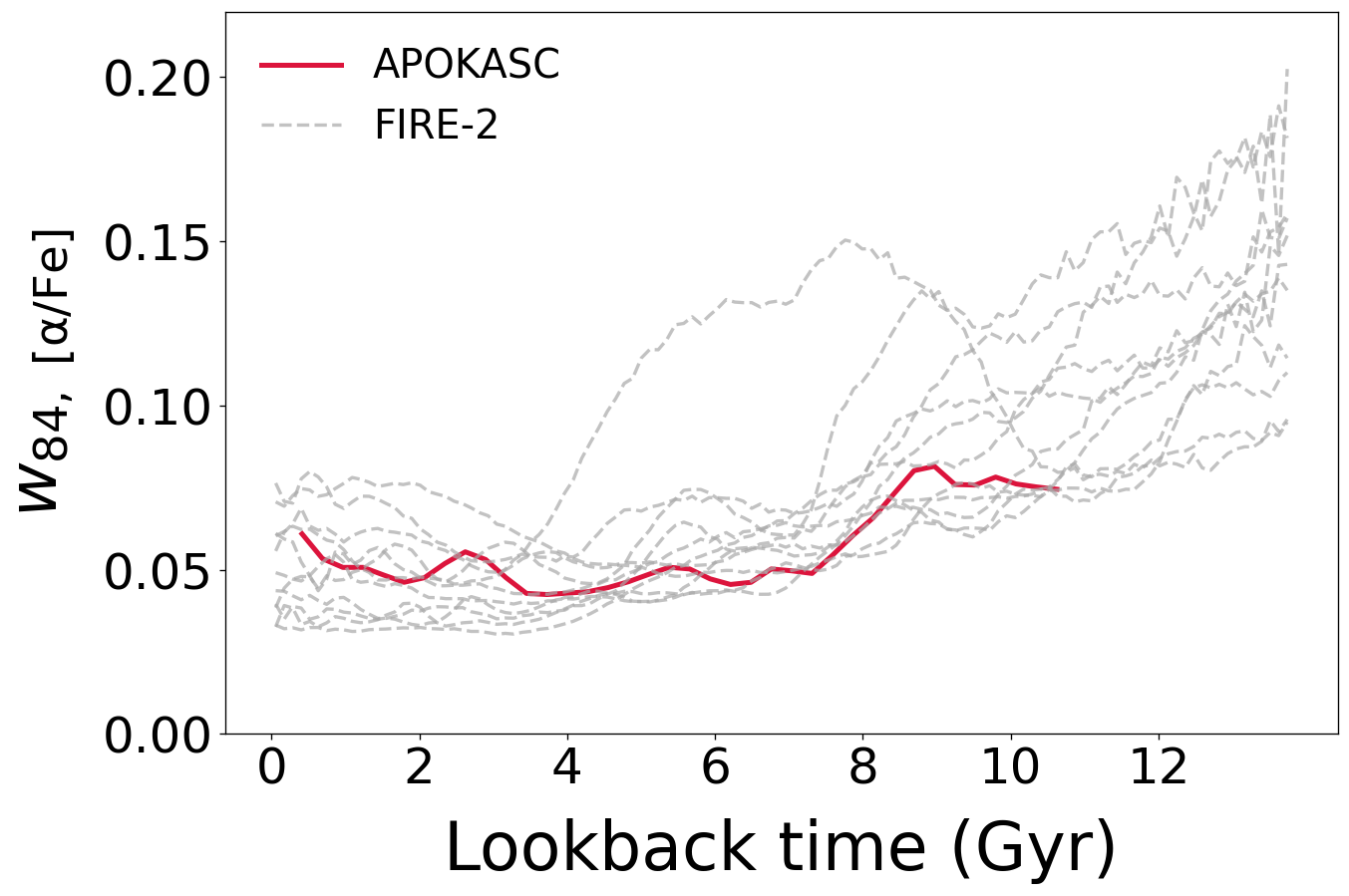}
    \caption{$w_{84, ~\mathrm{[\alpha/Fe]}}$ as a function of time for stars from the APOKASC-2 catalog. The decline around 8 Gyrs ago marks the end of the bursty epoch of star formation in the Milky Way. Beyond ~10 Gyr, the measurement becomes compromised by poor statistics. $w_{84}$, measured for the FIRE-2 galaxies with observational uncertainties close to ones in APOKASC-2 ($\sigma_{\tau} / \tau = 0.1$ and $\sigma_{\mathrm{[O/Fe]}} = \sigma_{\mathrm{[Fe/H]}} = 0.01$ dex), are overplotted in grey.}
    \label{fig:apogee}
\end{figure}

We measured the width of the $\mathrm{[\alpha/Fe]}$ distribution as a function of time for metal-rich (above 75th percentile by [M/H] at a given age bin) stars. Figure \ref{fig:apogee} shows the evolution of $w_{84, ~\mathrm{[\alpha/Fe]}}$ measured for stars from the APOKASC-2 catalog (red line) along with $w_{84}$ for simulated galaxies (grey lines). The error model applied to the simulated galaxies includes age and abundance uncertainties that resemble uncertainties in observational data. For the APOKASC-2 we show the measurement only up to 11 Gyr due to the lack of old stars however that does not affect our estimate of the transition time. The overall behavior of $w_{84}$ for APOKASC-2 is similar to $w_{84}$ in the simulations from FIRE-2 --- high value of $w_{84}$ at early times followed by a decline and transition to a plateau. The visual inspection of the observational $w_{84}$ suggests that the bursty phase of star formation in the Milky Way ended 7-8 Gyr ago.

\section{Discussion}
\label{sec:discussion}
For the case of simulated Milky-Way-like galaxies, we have demonstrated that the scatter of the [O/Fe] distribution measured for metal-rich stars at a given age bin is correlated with the burstiness of historical star formation, which allows an estimation of the transition time from bursty to steady regime. We have also tested whether this result could be achieved with the measurements of the scatter for the [Fe/H] and [O/H] distributions and found that $w_{84,\mathrm{[Fe/H]}}$ and $w_{84,\mathrm{[O/H]}}$ do not provide as good proxy for burstiness as $w_{84, \mathrm{[O/Fe]}}$, especially for the galaxies from Local Group-like pairs. 
Our findings are in agreement with \citet{Carillo22} where they found that for the simulation \texttt{m12i} an increased scatter around the age-[X/Fe] trend at earlier times reflects the bursty mode of star formation. 

For the Milky Way, the visual inspection of the evolution of $w_{84, \mathrm{[\alpha/Fe]}}$ suggests that the bursty phase in our galaxy ended 7-8 Gyrs ago. A few recent works explored the connection between the morphology of the FIRE-2 galaxies and the mode of their star formation. \citet{Yu} showed that during the bursty mode stars were born with thick-disk kinematics and the formation of the thin disk correlates with the transition to the steady regime; they also estimated the end of the bursty phase for the Milky Way to be $\sim$ 6.5 Gyr ago based on a comparison between the Milky Way and one of the FIRE-2 simulated galaxies, \texttt{Romeo}, which is later than our estimate. \citet{Gurvich22} analyzed the physics of gas disc settling and demonstrated that the rapid development of the rotationally-supported gas disk coincides with the end of the bursty phase. 
Therefore we can expect that the transition time in the Milky Way might match the onset of thin disk formation. 
For example, \citet{Snaith2014} derived the Milky Way's star formation history from fitting the age-[Si/Fe] relation with a chemical evolution model and suggested the formation epoch of the thick disk to be 13-9 Gyr ago, followed by a $\sim 1$ Gyr dip in star formation and then the formation of the thin disk for the last $\sim 8$ Gyr, which is in agreement with our estimates. \citet{Xiang&Rix} used the data from Gaia EDR3 and LAMOST to study the stellar age-metallicity distribution $p(\tau, \mathrm{[Fe/H]})$ and also found that the distribution splits at $\sim 8$ Gyr with the younger part corresponding to the formation of the thin disk; 
\citet{Conroy22} explored the chemistry, ages and kinematics of high-$\alpha$ in-situ stars and proposed that the oldest metal-poor in-situ stars were born during the ``simmering phase'' characterized by low star formation efficiency, which was followed by ``boiling phase'' when the efficiency of star formation has increased dramatically. Their estimate of the end of high-$\alpha$ disk formation ($z\approx 1$) also agrees with our estimates for the end of bursty phase.
Early (7-8 Gyrs) transition to steady star formation and, therefore, early end of the thick disk formation is also in qualitative agreement with the results of several chemical evolution models that predict early and rapid formation of halo and thick disk and more prolonged thin disk formation \citep{chiappini_2im2, micali_3im}.
A similar picture of the chaotic and isotropic early phase and subsequent rapid disk formation emerges from the analysis of kinematics and abundance data of in-situ stars in \citet{BelokurovKravtsov}, who also show the larger scatter in elemental abundances for the low-metallicity (i.e. older) stars and emphasize that the scatter increases when Al, Si, N and O ratios to either Fe or Mg are considered. To summarize, our estimate of the transition time to the steady phase is in line with other estimates of the end of the thick disk formation that are based on completely different methods.

One of the still remaining questions is whether the chevron shape of the [O/Fe]-[Fe/H] distribution and associated elevated level of scatter of alpha-elements abundances during the bursty phase is related to numerical effects in the simulation. 
The effect of implemented stellar feedback can be qualitatively understood in a following way: stronger feedback results in stronger bursts and faster depletion of star-forming gas reservoirs (see e.g. Fig.11 in \citet{Sparre}). This should lead to a more pronounced upper arm of the chevron and, therefore, increased scatter. For the case of stronger bursts we also expect them to be more isolated and scattered temporally, similar to the bursts in dwarf galaxies \citep{patel2021}.  

Insufficient metal mixing in simulations can artificially increase [O/Fe] scatter. In the FIRE-2 simulations we analyzed in this paper, metal mixing is explicitly modeled to account for unresolved (or subgrid) turbulent eddies and metal diffusion between boundaries of neighbouring resolution elements. \citet{Hopkins_FIRE2} showed that this numerical diffusion term has negligible effect on star formation histories; however it is important for proper modeling of chemical evolution and abundance ratios. This was investigated in \citet{Escala} for dwarf galaxies in FIRE-2. They found that including turbulent metal diffusion tends to reduce the intrinsic scatter in [$\alpha$/Fe] vs [Fe/H] and allows agreement between the width of the metal distribution function and dispersion of $\alpha$-element abundance ratios of the simulated and observed galaxies. Stronger mixing would lead to a further reduction of the chevron and tightening of the [O/Fe] - age relation. A different implementation of metal mixing in the simulations might therefore affect the magnitude of the effect we have identified, but is unlikely to change its qualitative existence.

\citet{Escala} compared intrinsic scatter in [$\alpha$/Fe] vs [Fe/H] and its evolution between simulated and real dwarf galaxies. Due to the correlation of [Fe/H] with age they conclude that intrinsic scatter of [$\alpha$/Fe] at fixed [Fe/H] might serve as a characteristic of ISM homogeneity. For the galaxies in their sample they found that evolution of [Si/Fe] scatter is characterised by presence of individual peaks resulting from starbursts which is in agreement with our finding for more massive galaxies. Outside of the peaks, their reported scatter stays near zero through all the time which is interpreted as a sign of well-mixed ISM at almost all times. In contrast to \citet{Escala} our measurement of scatter is nowhere near zero during the bursty phase. Part of the explanation is that, unlike in dwarfs, in Milky Way-like galaxies bursts could affect the abundance distribution only in parts of the galaxy. Also, bursts in massive galaxies are more frequent than in dwarfs, therefore in our analysis each considered time bin (230 Myr) covers a few burst episodes.  
In other words, instead of focusing on separate starbursts we study the evolution of the star formation regime on a longer timescale. The assumption of a well-mixed ISM during bursty periods needs some clarification: galactic disks during bursty epochs indeed lack the radial gradients that exist in present-day disks \citep{Bellardini}; however, formation of two distinct arms on the [Fe/H]-[O/Fe] plane requires the ISM to have clumps which might have different elemental abundance ratios than that of the smooth ISM around them, and therefore abundances in the disk experience variations on smaller scales. That is confirmed by the work of \citet{Bellardini}, who showed that azimuthal scatter of abundances in gas disks increase at higher redshift.  

In our analysis we included only stars formed within the central 20 kpc of the host, therefore we mostly ignored the effect of mergers. Accreted stars could be filtered out based on the analysis of their chemodynamics so they would not affect the computation of $w_{84}$. However mergers could induce strong bursts of star formation \citep{sparre_mergers} and lead to a potential mismatch between burstiness and scatter of [O/Fe] as in case of \texttt{m12b} or elevated scatter in the steady regime as in the case of the recent merger in \texttt{m12f}. 
Therefore a merger event occuring near the end of the bursty phase might affect the estimation of the transition time. 
We tested the effect of including ex-situ stars by computing transition times for all stars within the central 100 kpc of the host at $z=0$ and found the effect negligible for almost all galaxies except for \texttt{m12c} where the difference was around 260 Myr, which is less than the typical uncertainty of our method.

The scatter of the alpha abundances in the Milky Way shows the same behavior as in the FIRE-2 simulated galaxies, for which the scatter and star formation burstiness are correlated. We have suggested that we can therefore interpret the abundances of Milky Way stars to infer the history of its star formation character. However it would be interesting to explore evolution of alpha abundance distribution in galaxies with wider range of star formation histories, including those simulations that do not show early bursty period e.g \citet{Park}.

\section{Conclusion}
\label{sec:conclusion}
For 11 Milky Way-mass galaxies from the FIRE-2 simulations we analysed the distribution of [O/Fe] as a function of time for high-metallicity stars formed in a host galaxy. We found that the width of the alpha element abundance ratio distribution, $w_{84}$, is correlated with burstiness of star formation, $\sigma_{10}/{\left<\mathrm{SFR_{500}}\right>}$
and this correlation is robust to observationally motivated age and abundance uncertainties and selection function. 
The wider distribution of alpha-elements during bursty periods is related to the behavior of the stellar distribution on the [Fe/H]-[O/Fe] plane --- an intense burst of star formation followed by depletion of the star-forming gas reservoir leads to the large amount of core collapse supernovae enriching the ISM mainly with alpha-elements, while iron mostly comes from the Type Ia supernovae.
For sufficiently intense bursts, this happens faster than the mixing timescale, resulting in the formation of a ``sideways chevron'' shape on the [Fe/H]-[O/Fe] plane. Measuring the scatter of [O/Fe] for stars above 75th percentile in [Fe/H] at a given time captures the presence of chevron arms. In the steady regime of star formation, mixing between different enrichment sites occurs more effectively and leads to a more uniform distribution of elements in the disk at a given radius. 
We estimate the end of the bursty phase as the time when $w_{84}$ as measured from data convolved with different age uncertainties converge. We find that this transition time is in excellent agreement with that derived directly from the star formation history. 
We applied this to the Milky Way and computed $w_{84}$ for data from the Second APOKASC Catalog to estimate the end of the bursty star formation mode. We found that $w_{84}$ computed from the observational dataset shows similar behavior, and implies that the Milky Way transitioned from its bursty phase to its steady phase $\approx 7-8$ Gyr ago, earlier than simulations.

\section*{Acknowledgements} 
We thank the reviewer for providing us with constructive comments. We are also grateful for useful discussions with Jamie Tayar, Sanjib Sharma, and Eliot Quataert.

Simulations used in this paper were run on the Caltech computer
cluster “Wheeler,” allocations AST21010 and AST20016 supported by the NSF and TACC. JB was supported by HST grant AR-12837. AW received support from: NSF via CAREER award AST-2045928 and grant AST-2107772; NASA ATP grant 80NSSC20K0513; HST grants AR-15809, GO-15902, GO-16273 from STScI. CAFG was supported by NSF through grants AST-1517491, AST-1715216, and CAREER award AST-1652522; by NASA through grant 17-ATP17-0067; and by a Cottrell Scholar Award from the Research Corporation for Science Advancement.

\section*{Data availability}
The FIRE-2 simulations are publicly available \citep[][]{wetzel2022} at \url{http://flathub.
flatironinstitute.org/fire}. Additional FIRE simulation
data is available at \url{https://fire.northwestern.edu/data}.
A public version of the GIZMO code is available at \url{http:
//www.tapir.caltech.edu/~phopkins/Site/GIZMO.html}.




\bibliographystyle{mnras}
\bibliography{example} 




\appendix

\section{Burstiness and abundance patterns for all simulations}
\label{sec:appA}

\begin{figure*}
	\includegraphics[width=2\columnwidth]{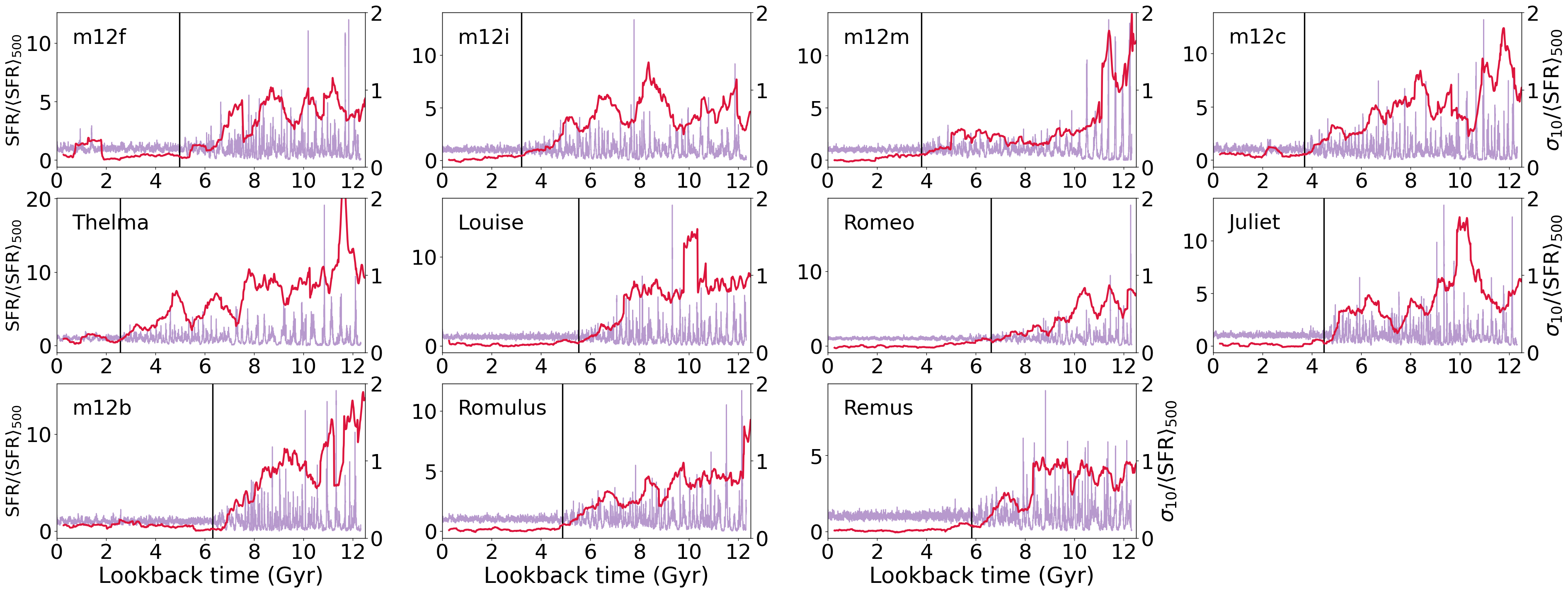}
    \caption{Star formation histories measured in 1 Myr bins using the ages of stars at z = 0 (purple line) of all galaxies in our sample. All galaxies at high redshifts show a bursty mode of star formation, characterized by rapid variations with high amplitude, which is followed by steady star formation at more recent times (past $\sim 5$ Gyr). The red line shows burstiness, $\sigma_{10}/\mathrm{\left<SFR\right>_{500}}$ and vertical lines mark transition times, defined as the time when burstiness falls below 0.2.}
    \label{fig:std_all}
\end{figure*}

Each panel of Fig.~\ref{fig:std_all} shows the evolution of the instantaneous star formation rate (SFR), measured in 1 Myr bins and normalized by the SFR averaged over 500 Myr rolling window (purple line). At early times ($\displaystyle t_\mathrm{lookback} \gtrsim$ 5-6 Gyr) star formation was ``bursty'' and characterized by high variability with 10-40 Myr timescale. Recently, over the past 4-6 Gyr, star formation proceeded in a ``steady'' mode with relatively constant star formation rate with the exception of a few isolated bursts like in \texttt{m12f} which are related to minor mergers. The red lines show burstiness, $\sigma_{10}/\mathrm{\left<SFR\right>_{500}}$ and vertical lines mark transition times, defined as the time when burstiness falls below 0.2.

\begin{figure*}
	\includegraphics[width=2\columnwidth]{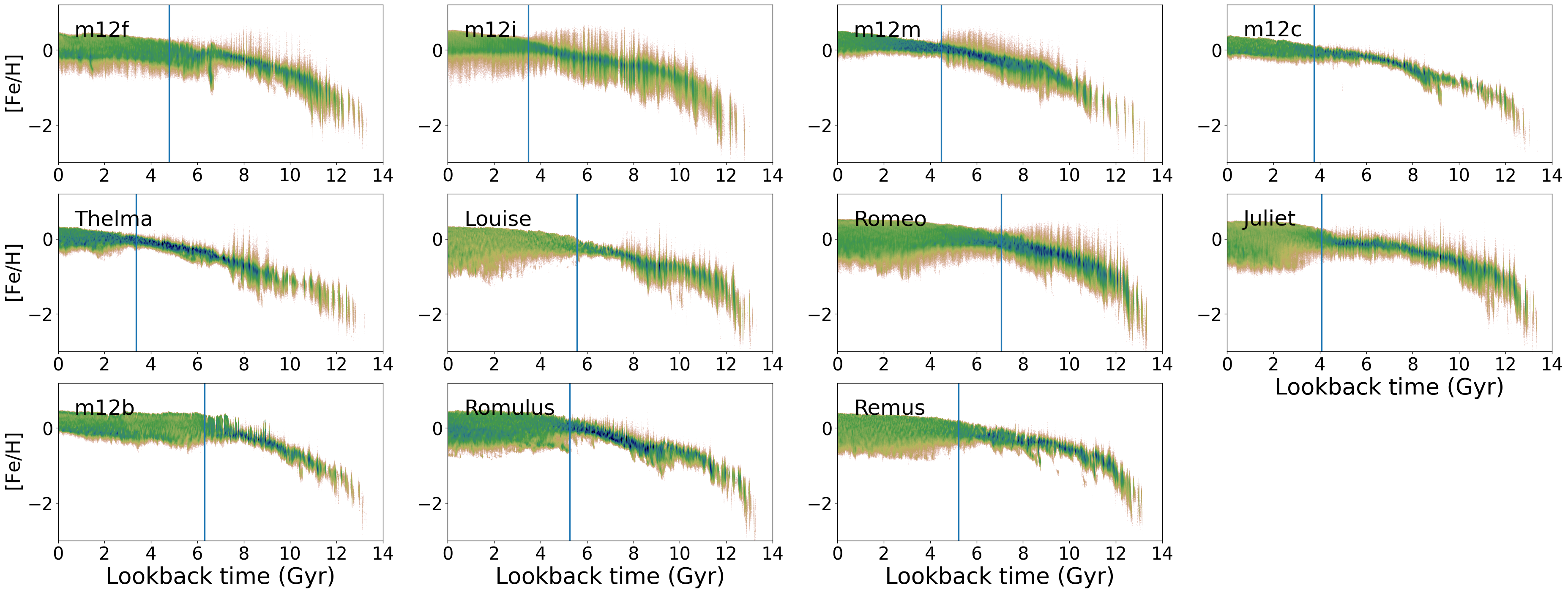}
    \caption{Evolution of [Fe/H] with lookback time. The vertical line marks the end of the bursty phase as on the corresponding panel of Fig.\ref{fig:std_all}.}
    \label{fig:age_feh_all}
\end{figure*}

\begin{figure*}
	\includegraphics[width=2\columnwidth]{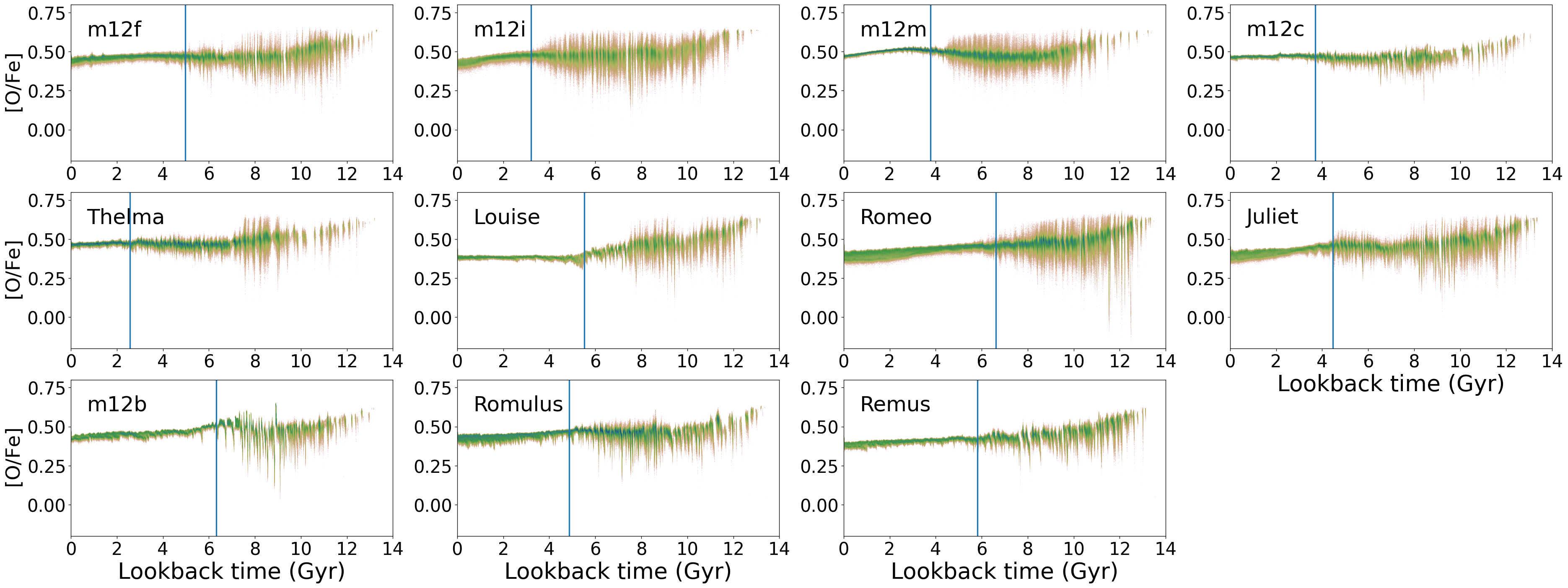}
    \caption{Age-[O/Fe] distribution for all 11 galaxies plotted for stars with metallicity higher than the 75th percentile value at a given time bin.}
    \label{fig:age_ofe_all}
\end{figure*}

Fig.~\ref{fig:age_feh_all} shows iron abundance [Fe/H] and Fig.~\ref{fig:age_ofe_all}  shows relative oxygen abundance [O/Fe] as a function of lookback time for all galaxies in the sample with vertical lines marking transition times derived from analysis of star formation history. Each galaxy demonstrates two
distinct modes of enrichment with oxygen and iron. The scatter of the [O/Fe] distribution rapidly decrease after the end of the bursty phase.  

\begin{figure*}
	\includegraphics[width=2\columnwidth]{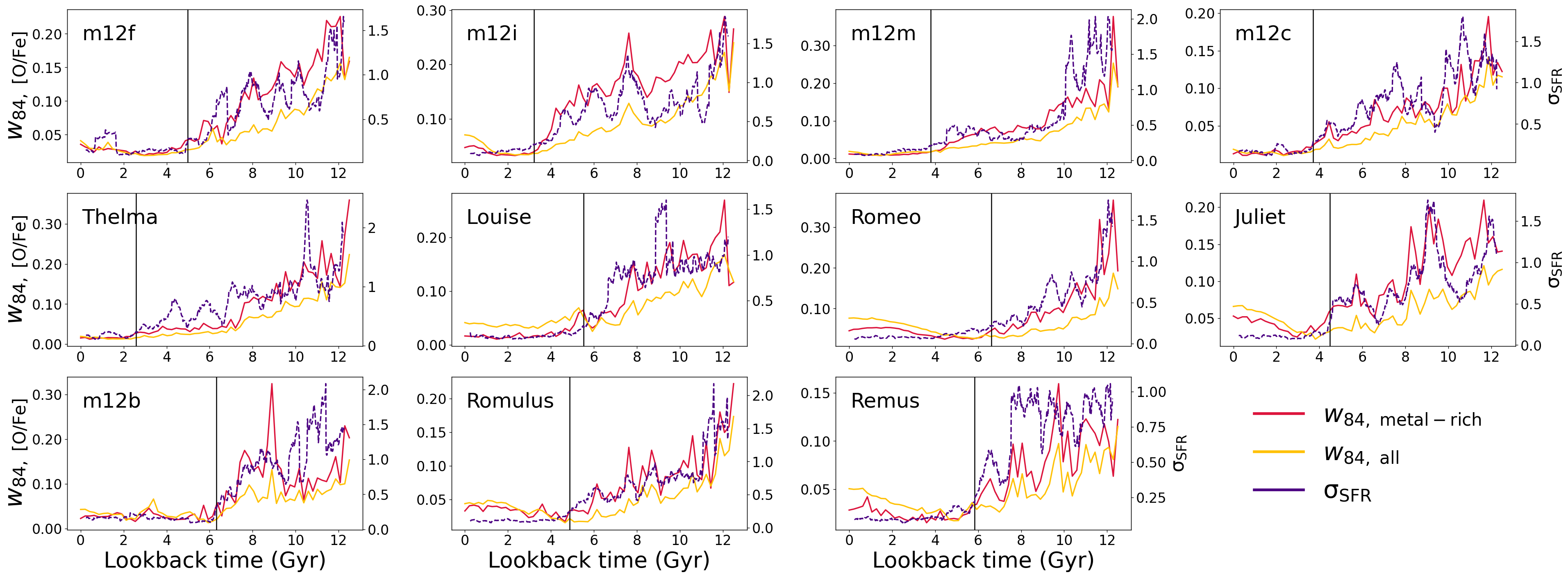}
    \caption{Burstiness, $\sigma_{10}/\mathrm{\left<SFR\right>_{500}}$ (blue curve), and the width of the [O/Fe] distribution, $w_{84}$, as a function of time, measured for stars above the 75th percentile of [Fe/H] at a given age (red curve), vs. for all stars (yellow curve). The scatter is well correlated with burstiness, however $w_{84}$ measured for metal-rich stars (red line) shows higher contrast between values measured in bursty and steady modes.}
    \label{fig:compare_high_fe_all}
\end{figure*}

Fig.~\ref{fig:compare_high_fe_all} shows burstiness, $\sigma_{10}/\mathrm{\left<SFR\right>_{500}}$ (blue curve), and the width of the [O/Fe] distribution, $w_{84}$, as a function of time, measured for stars above the 75th percentile of [Fe/H] at a given age (red curve), vs. for all stars (yellow curve). For all galaxies the scatter is well correlated with burstiness, however $w_{84}$ measured for metal-rich stars (red line) shows higher contrast between values measured in bursty and steady modes.

\begin{figure*}
	\includegraphics[width=2\columnwidth]{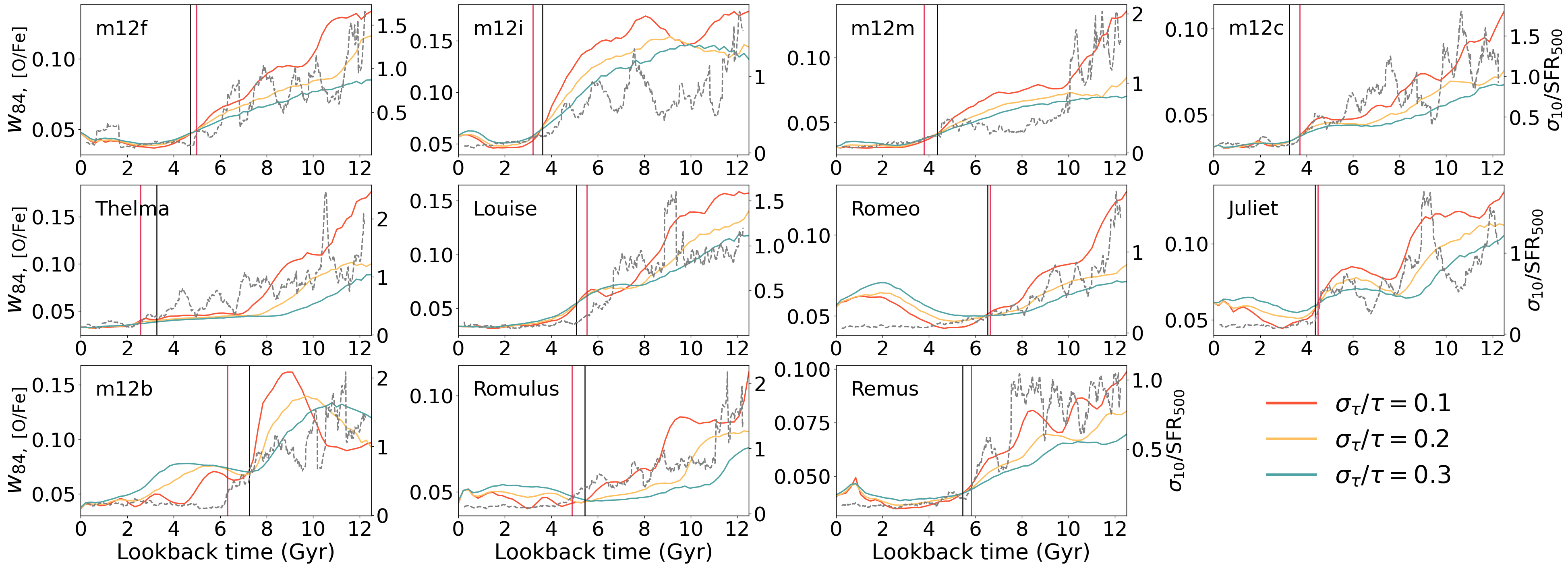}
    \caption{Each panel shows the [O/Fe] scatter, $w_{84}$, versus time for a different simulated galaxy, computed using three different age error models and abundance uncertainty of 0.01 dex. The time of convergence of three $w_{84}$ functions (vertical black line) is a close match to the transition time to the steady regime marked by the red vertical line. Burstiness is overplotted in gray to emphasize the agreement between the scatter and mode of star formation.}
    \label{fig:burstiness_w84_all}
\end{figure*}

Fig.~\ref{fig:burstiness_w84_all} shows the [O/Fe] scatter, $w_{84}$, versus time, computed using three different age ($\tau$) error models ($\sigma_{\tau}/\tau$ = {0.1, 0.2, 0.3}) and abundance uncertainty of 0.01 dex. The time of convergence of three $w_{84}$ functions (vertical black line) is a close match to the transition time to the steady regime marked by the red vertical line. Burstiness is overplotted in gray to emphasize the agreement between the scatter and mode of star formation.

\section{Case of \texttt{m12b}}
\label{m12b}
\begin{figure*}
	\includegraphics[width=\columnwidth]{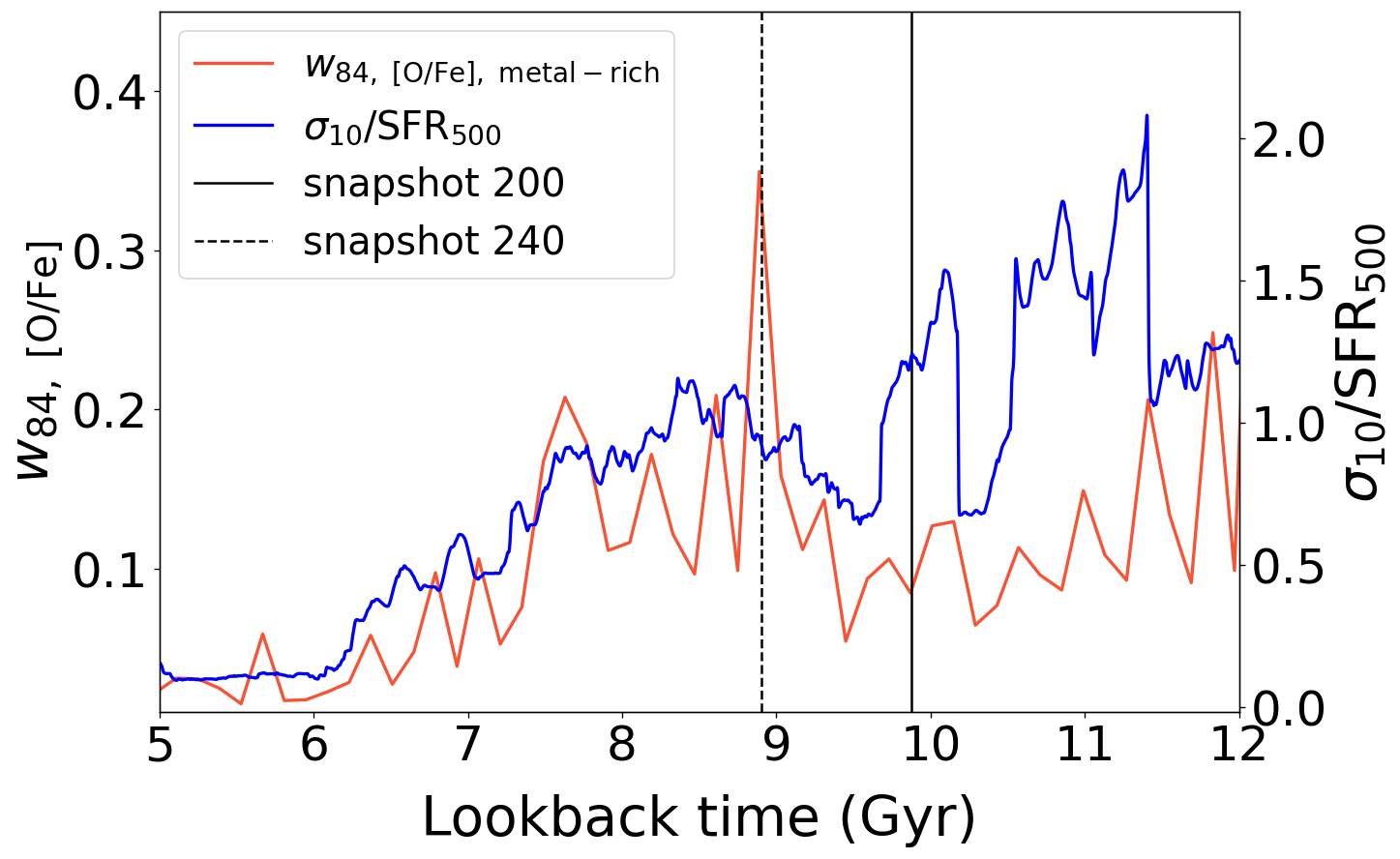}
    \caption{Burstiness and [O/Fe] scatter for galaxy \texttt{m12b} with vertical lines indicating the times of snapshots 200 and 240. The decrease of the burstiness has a more monotonical behaviour, while $w_{84}$ shows a very high-amplitude peak at 9 Gyr which does not seem to match any particularly strong individual burst episode.}
    \label{fig:m12b_burst}
\end{figure*}

\begin{figure*}
    \centering
    \includegraphics[width=2.\columnwidth]{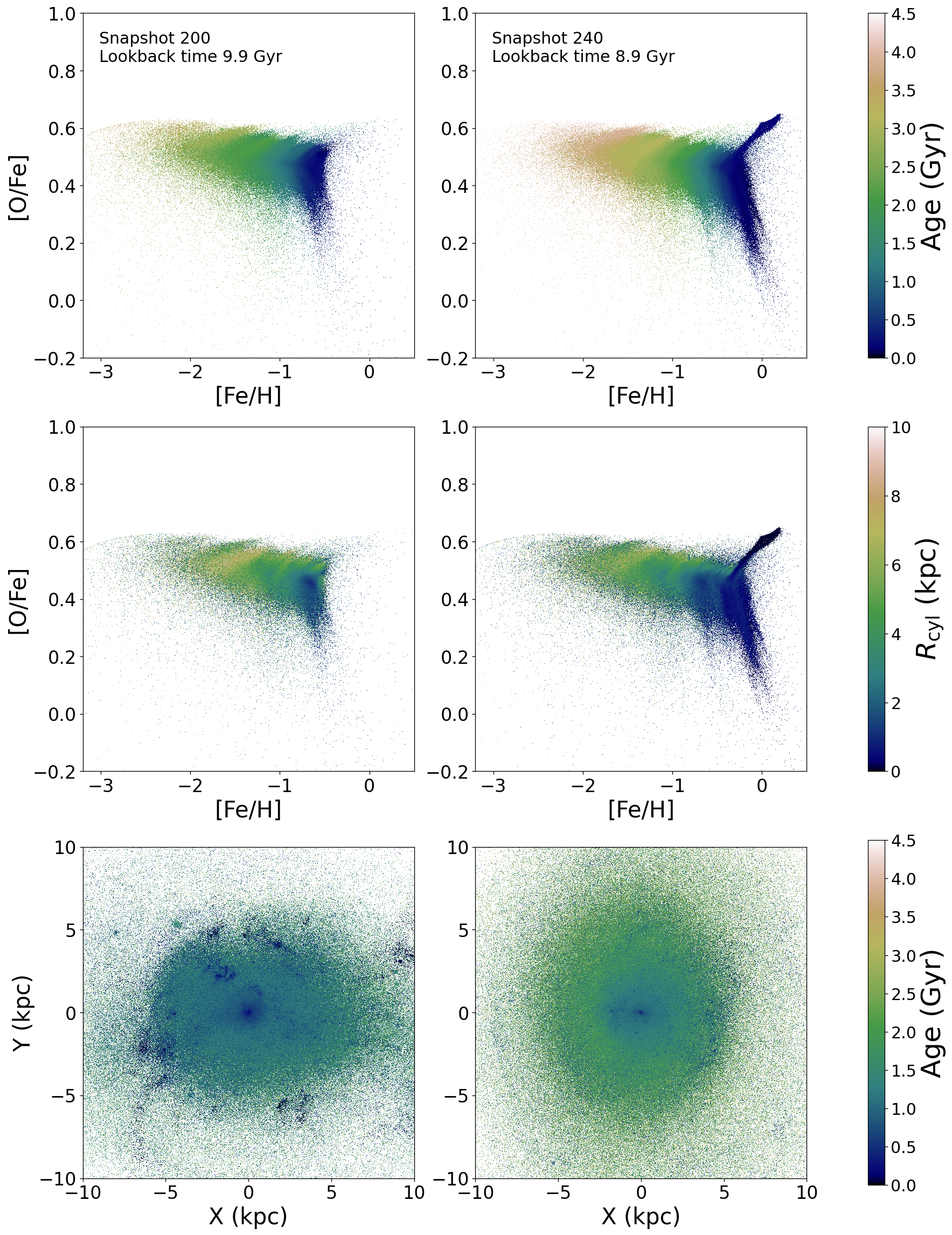}
    \caption{\textit{Top row}---[Fe/H]-[O/Fe] diagrams for galaxy \texttt{m12b} at snapshots 200 \textit{(left)} and 240 \textit{(right)} coloured by the age of the stars at the time of the snapshot, \textit{middle row}--- same as the top row but coloured by cylindrical radius, \textit{bottom row}--- distribution of stars on the XY plane. The earlier snapshot shows the presence of multiple clumps of ongoing star formation, while the galaxy at the later snapshot has all its star formation confined in the compact central region which gives a rise to the extended upper ``arm'' on the corresponding [Fe/H]-[O/Fe] diagram.}
    \label{fig:m12b}
\end{figure*}

Despite a general good agreement between the burstiness and the scatter of [O/Fe], few galaxies show some mismatches between the $\sigma_{10}/\mathrm{\left<SFR\right>_{500}}$. One such striking example is the simulation \texttt{m12b}, which have the peaks of burstiness and $\sigma_{\mathrm{[O/Fe]}}$ separated by $\sim$2 Gyr at high redshifts. To investigate the source of this discrepancy we compared the stellar distribution at snapshots 200 and 240 corresponding to lookback times 9.7 Gyr and 8.7 Gyr (Fig.~\ref{fig:m12b_burst}). Both snapshots capture the galaxy in a bursty stage, so $\sigma_{\mathrm{SFR}}$ and $w_{84}$ are both relatively high; however, while burstiness declines on the interval bounded by these snapshots, the scatter of [O/Fe] rises dramatically. Both snapshots 200 and 240 have a chevron shape, clearly seen on the [Fe/H]-[O/Fe] diagram colour-coded by the age of the stars (top left panel on Fig.~\ref{fig:m12b}); however the distribution for the later snapshot has a striking feature --- a very long upper arm corresponding to a short and intense burst of star formation which occured in a very compact (less than 0.5 kpc) region near the center. 

This example supports the idea that the chevron shape and associated increased scatter in [O/Fe] as a function of age are related to episodes when intense star formation is confined to a number of dense clumps. The clumpy character of star formation in high-redshift galaxies has been confirmed in observations \citep[e.g.][]{clumps_obs} and in FIRE simulations \citep{Oklopcic}.
In \texttt{m12b}, the stellar distribution at snapshot 200 contains multiple concentrated clumps with very recent star formation. Snapshot 240, which is 1 Gyr later, has an overall smoother stellar distribution, punctuated by a small but intense center of star formation, which rapidly produces new stars and deposits a large amount of alpha-elements into the ISM on timescales shorter than delay of supernovae Type Ia. This is the source of the arm on the [Fe/H]-[O/Fe] diagram. \texttt{m12b} had a merger with a satellite in between snapshots 200 and 240; this merger might be a cause of the central burst of star formation given that galaxy interactions lead to fuel-driven enhancement of star formation in the central kiloparsec, as discussed in \citet{Moreno2021}.

To sum up, the case of \texttt{m12b} with the decoupling between $w_{\mathrm{84}}$ and $\sigma_{\mathrm{SFR}}$ demonstrates that dispersion of [O/Fe], $w_{84}$, is very sensitive to spatially concentrated intense bursts of star formation even in cases where those bursts are not dominating the overall star formation mode of the galaxy.

\bsp	
\label{lastpage}
\end{document}